\documentclass{article}

\usepackage[english]{babel}

\usepackage[letterpaper,top=2cm,bottom=2cm,left=3cm,right=3cm,marginparwidth=1.75cm]{geometry}

\usepackage{amsmath,caption,indentfirst}
\usepackage{graphicx}
\usepackage[colorlinks=true, allcolors=blue]{hyperref}
\usepackage{amsmath}
\title{Glymphatic Clearance in the Optic Nerve: A Multidomain Electro-osmostic Model}
\author{Shanfeng Xiao \thanks{The Department of Mathematics, Changzhi University, 73 Baoningmen East street, Changzhi 046011, Shanxi Province, China}
	\and Huaxiong Huang \thanks{
    Zu Chongzhi Center, Duke Kunshan University, 8 Duke Ave, Kunshan, China;
    Laboratory of Mathematics and Complex Systems, MOE, Beijing Normal University,  Beijing 100875, China;
		Department of Mathematics and Statistics, York University, Toronto, ON M3J 1P3, Canada}
	\and Robert  Eisenberg \thanks{Department of Applied Mathematics, Illinois Institute of Technology, Chicago, IL 60616, USA}
	\and  Zilong Song \thanks{Math and Statistics Department, Utah State University, Old Main Hill, Logan, UT 84322, USA}
	\and Shixin Xu \thanks{Zu Chongzhi Center, Duke Kunshan University, 8 Duke Ave, Kunshan, China}
}

\begin{document}
	\maketitle
 

	\begin{abstract}
Effective metabolic waste clearance  and maintaining ionic homeostasis are essential for the health and normal function of the central nervous system (CNS). To understand its mechanism and the role of fluid flow, we develop a multidomain electro–osmotic model of optic–nerve microcirculation (as a part of the CNS) that couples hydrostatic and osmotic fluid transport with electro–diffusive solute movement across axons, glia, the extracellular space (ECS), and arterial/venous/capillary perivascular spaces (PVS). Cerebrospinal fluid enters the optic nerve via the arterial parivascular space (PVS-A), passes both the glial and ECS before exiting through the venous parivascular space (PVS-V). Exchanges across astrocytic endfeet are essential and they occur in two distinct and coupled paths: through AQP4 on glial membranes and gaps between glial endfeet, thus establishing a mechanistic substrate for two modes of glymphatic transport, at rest and during stimulus–evoked perturbations. Parameter sweeps show that lowering AQP4–mediated fluid permeability or PVS permeability elevates pressure, suppresses radial exchange (due mainly to hydrostatic pressure difference at the lateral surface and the center of the optic nerve), and slows clearance, effects most pronounced for solutes reliant on PVS–V export. The model reproduces baseline and stimulus–evoked flow and demonstrates that  PVS–mediated export is the primary clearance route  for both small and moderate solutes. Small molecules (e.g., A$\beta$) clear faster because rapid ECS diffusion broadens their distribution and enhances ECS–PVS exchange, whereas moderate species (e.g., tau monomers/oligomers) have low ECS diffusivity, depend on trans–endfoot transfer, and clear more slowly via PVS–V convection. 
Our framework can also be used to explain the sleep–wake effect mechanistically: enlarging ECS volume (as occurs in sleep) or permeability increases trans–interface flux and accelerates waste removal. Together, these results provide a unified physical picture of glymphatic transport in the optic nerve, yield testable predictions for how AQP4 function, PVS patency, and sleep modulate size–dependent clearance, and offer guidance for targeting impaired waste removal in neurological disease.
	\end{abstract}
	
	\section{Introduction}
	
	 The glymphatic system, which consists of glial cells and perivascular space, is believed to play a crucial role in the clearance of metabolic waste from the central nervous system (CNS) and  maintaining neural health \cite{2012AParavascular,2015TheGlymphatic,2016Glymphatic,2017Impairment,2020Glymphatic,2024Modeling}. In particular, the flow of cerebrospinal fluid (CSF) through the perivascular spaces (PVS) of the brain and optic nerve is essential for clearing waste products like amyloid-$\beta$, which have been linked to neurodegenerative diseases such as Alzheimer's and cerebral amyloid angiopathy \cite{2012AParavascular,2018Flow,2022Perivascular}. Recent studies have demonstrated that CSF enters the brain along arterial PVS and drives the clearance of solutes from the interstitial fluid (ISF) at downstream locations \cite{1992Alzheimer,2016Theamyloid} providing insight into the general 'blood brain barrier'. This glymphatic mechanism has also been implicated in the clearance of potassium, a process critical for maintaining ionic homeostasis in the CNS \cite{2020Glymphatic}. 
	
	Importantly, the glymphatic system exhibits a strong dependence on sleep. During sleep, the extracellular space (ECS) in the brain expands significantly, enhancing the efficiency of CSF flow and waste clearance \cite{2013sleep,fultz2019coupled,2024Neuronal}. This sleep-dependent enlargement of the ECS facilitates the convective movement of fluid through the perivascular spaces, allowing for more effective removal of metabolic waste products, including amyloid-$\beta$ and potassium ions. Disruptions in sleep have been shown to impair glymphatic function, potentially contributing to the accumulation of toxic metabolites and the progression of neurodegenerative diseases \cite{2013sleep,2020Glymphatic}. These findings underscore the critical role of sleep in maintaining CNS homeostasis and highlight the importance of understanding the interplay between sleep, ECS dynamics, and glymphatic function.
	
	Most of the current investigations focused on mechanical factors such as hydrostatic pressure gradient and astrocytic endfoot movement due to arterial pulsations and volume changes as the driving forces of fluid flow in this glymphatic system \cite{2016Glymphatic,2003Arterial}. 
	More recently,  machine learning (ML) techniques have also been applied to explore  the flow within the perivascular spaces \cite{boster2023artificial,chou2024machine}, with a detailed review available in \cite{2022glymphatic}. However, these new ML techniques do not provide mechanistic insights.  
	
	On the other hand, it is well known \cite{Feher2012Quantitative,silverthorn2019human} that osmosis is a major inescapable factor in generating fluid movement in biological tissues \cite{2020ATridomain}. Flow in biological systems is controlled by the structure and properties of membranes in organs and tissues as they modulate the unavoidable physical process of osmosis \cite{2023Structural}. It is, therefore, natural to ask whether osmotic force plays an important role in producing fluid flow in the glymphatic system. 
	    
    Our general perspective is that the location, density and type of ion channels, water channels, and active transport systems in the brain provide electro-osmotic glymphatic transport that is up and down regulated during sleep to clear waste from the brain. The biological structure modulates the physical forces of electro-osmosis that always exist and a prime example is the kidney. The lens of the eye is another tissue that depends on an electro-osmotic pump to maintain its biological function, as studied in great detail experimentally and theoretically, see references in \cite{1985Epithelial,1997Physiological,2019ABidomain,2011Lens}.
	
	This perspective on glymphatic flow needs to be tested before it is accepted as reality, but having an explicit working hypothesis makes testing easier to focus and more effective. It helps that we postulate a mechanism of electro-osmosis that occurs throughout animals and plants. A brain is a complex organ and many processes occur simultaneously, it is important and necessary to choose a model that is relatively simple but shares similar features of the brain. In particular,  we wish to explore the role played by osmosis in fluid flow and solute transport via the perivascular space as a well defined component of ``the blood brain barrier". 
	
	The optic nerve, part of the CNS, is structurally similar to the brain. It is surrounded by glial cells and includes narrow extracellular spaces \cite{2020Quantitative}. It consists of four primary regions: the intraocular nerve head, intraorbital region, intracanalicular region, and intracranial region \cite{2009Ischemic}. This study focuses on the intraorbital region, which is most of the optic nerve. The optic nerve is involved in waste clearance, with CSF entering via perivascular spaces around blood vessels and interacting with the glial cells that line these spaces \cite{2017Evidence}. The optic nerve forms an isolated system that can be studied experimentally as a physiological preparation \cite{1966Physiological,1966Effect}.
	
    Therefore, the optic nerve offers an ideal setting for understanding the dynamics of fluid flow and ionic transport in the central nervous system, as a simplified version of part of the ``blood brain barrier", under normal and pathological conditions that is accessible to experimental investigation. \cite{2017Aquaporin}.
	
	A previous study by Zhu et al. \cite{2020ATridomain} revealed that glial cells, particularly astrocytes, play a significant role in maintaining ionic homeostasis, namely potassium clearance in the optic nerve. During neural activity, potassium ions accumulate in the extracellular space, and their efficient clearance is necessary to maintain normal neuronal function and prevent excitotoxicity. The accumulation of potassium may play a role in signal processing on a time scale of hundreds of milliseconds to seconds.  Glial cells form an interconnected network via connexin-based gap junctions, which allows them to act as a syncytium, facilitating the redistribution and clearance of potassium from the extracellular space \cite{2017Astrocytic}. However, the role of perivascular space as a part of the glymphatic system was not considered. In this paper, we  extend our previous work and aim to explore the entire glymphatic system that consists of perivascular space and glial cells, which regulate fluid movement between the two through aquaporin-4 (AQP4) channels located in the glial endfeet facing the perivascular space \cite{2006Theimpact,2011Anaquaporin4}. Our model integrates key glial function and dynamics of the perivascular space, particularly their roles in coupled fluid transport and ion transport through convective, diffusive, and electrochemical transport mechanisms \cite{2019ABidomain}.
	
	Our model also considers the interaction between the optic nerve and its surrounding CSF, which flows through the perivascular spaces and subarachnoid space (SAS). This fluid serves as both a source of nutrients and a pathway for waste clearance \cite{2017Mechanisms}. CSF flows directly from the SAS into the   arterial perivascular spaces (PVS-A)  before flowing into the brain parenchyma via AQP4 channels, mixing with ECS, and then entering the  venuous perivascular spaces (PVS-V) for drainage via a convective flow. Fluid from the SAS can then drain into the meningeal lymphatic vessels (MLV) surrounding the superior sagittal sinus \cite{2020Quantitative}. Our model differentiates between the extracellular and perivascular spaces, and includes direct communication between CSF and  PVS-A. It enables use to explore the nature of interaction among glial cells, perivascular and extracellular spaces in greater details than previous models. 
	
By  carrying out a parametric study of the equilibrium state as well as a study of dynamics during neural firing, we show that the optic nerve's glymphatic system, mediated by glial cells and perivascular fluid dynamics, plays an essential role in maintaining ionic homeostasis via a mechanism of ion transport-fluid flow coupling. Our mathematical model provides a comprehensive framework to analyze this coupling processes, offering new insights into the interplay between fluid flow, ionic transport, and glial function in the optic nerve. Our model's adaptability allows for its application to diverse structures, including the brain with distinct channel and transporter distributions across the multiple compartments, extending its utility beyond the optic nerve.

	Our main finding is an intricate coupling among neural activities and various components of the glymphatic system: (i) when neurons are at rest, fluid flow in the glymphatic system is mainly in the radial direction, driven by the hydrostatic pressure difference (higher in PVS-A and lower in PVS-V); and (ii) when neurons fire, it induces a local perturbation to the ionic homeostasis and accumulation of potassium in ECS, which leads to changes in glial membrane potential and ionic fluxes flow, creates an osmotic pressure that drives fluid into the glia from ECS, which in turn causes fluid exchange between glial and perivascular space in the endfoot regions. The direction of the fluid, that exits PVS-A and enters PVS-V from glia, enhances fluid flow in the glymphatic system. In addition, the glia endfeet volume expands near PVS-A and shrinks near PVS-V due to the fluid extering glia from PVS-A and exiting glia to PVS-V, which reduces/increases endfoot gaps in PVS-A/PVS-V regions, reduces ECS to PVS-A fluid flow while enhances ECS-PVS-V fluid flow. 
		
	The remainder of this article is structured as follows: Section 2 presents the mathematical model for microcirculation in the optic nerve. Section 3 explores fluid dynamics in the resting state and under stimulus-induced conditions. In Section 4, we conduct a parametric study on permeability variations, examining how pathophysiological changes affect glymphatic function. Section 5 investigates the role of the glymphatic system in metabolic waste clearance, highlighting how glymphatic system enhances clearance efficiency. Finally, Section 6 provides concluding remarks,  including the limitations and future directions.

	\begin{figure}
		\centering
		\includegraphics[width=0.8\linewidth]{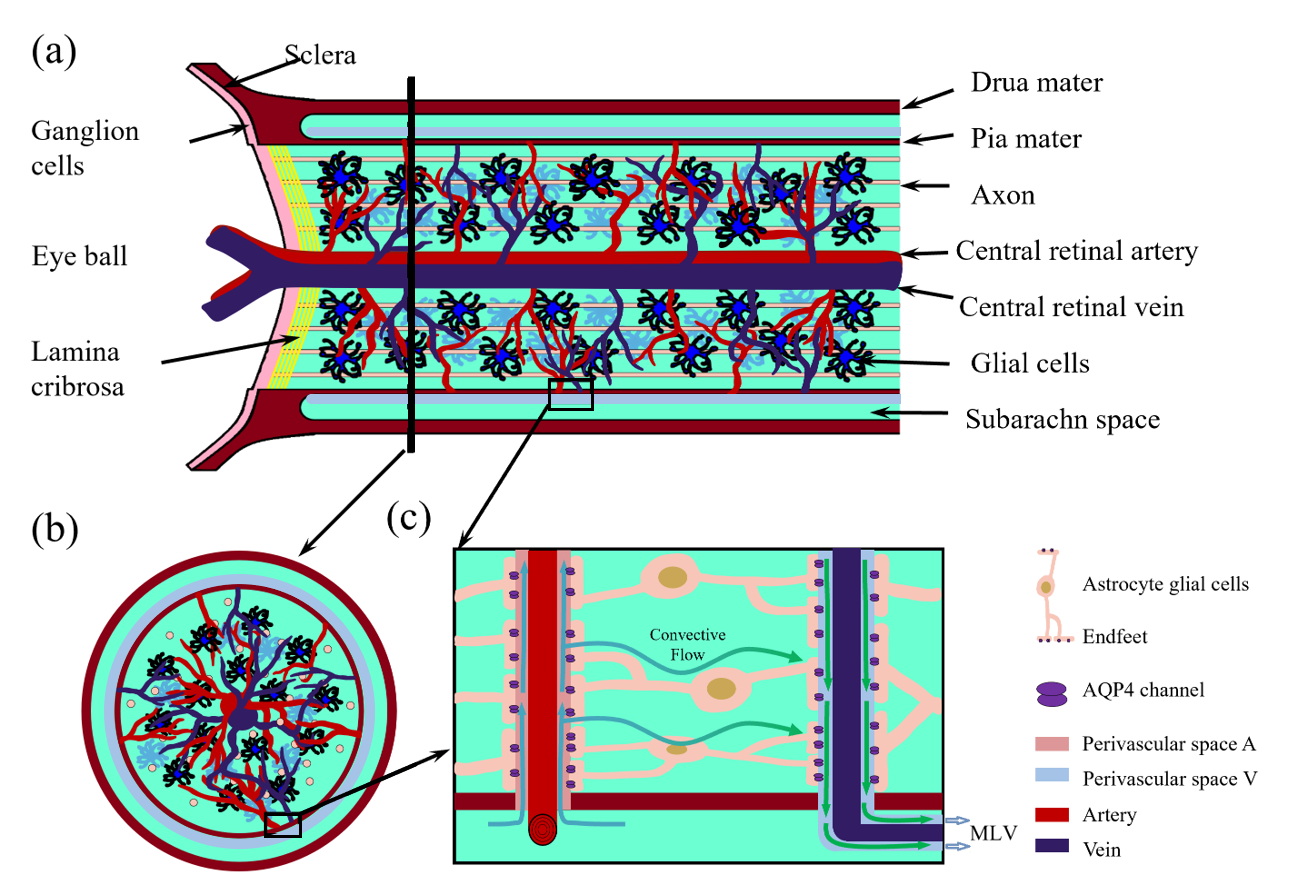}
		\caption{\label{fig:Optic-nerve}Optic nerve structure. (a) Longitudinal section of the optic nerve; (b) cross section of the optic nerve.}
	\end{figure}

	\section{Mathematical Model of Microcirculation of the Optic Nerve}
	
	\begin{table}
		\centering
		\caption*{\textbf{GLOSSARY}}
		\scalebox{0.7}{
			\begin{tabular}{ll}
				$C_{l}^{i}$: Ion $i$ concentration in compartment $l$, & $k_{e}^{l}$: Electroosmotic in compartment $l$, \\
				$\phi_{l}$: Electric potential in compartment $l$, & $\mathcal{M}_{a,b}$: Interface area form compartment $a$ to $b$ in per unit control volume, \\
				$p_{l}$: Hydrostatic pressure in compartment $l$, & $\kappa_{l}$: fluid permeability in compartment $l$, \\
				$\eta_{l}$:  Volume fraction in compartment $l$, & $\mu$: Fluid viscosity, \\
				$O_{l}$: Osmotic concentration in compartment $l$, & $L_{a,b}$: Hydrostatic permeability of interface form $a$ to $b$, \\
				$\mathbf{u}_{l}$: Fluid velocity inside of compartment $l$, & $g_{a,b}^{i}$: Conductance of interface form compartment $a$ to $b$ for ion $i$, \\
				$\mathbf{j}_{l}$: Ion $i$ flux inside of compartment $l$, & $\bar{g}^{i}$: Maximum conductance of axon membrane for ion $i$, \\
				$U_{a,b}$: Fluid flux across the interface form  compartment $a$ to $b$, & $g_{leak}^{i}$: Leak conductance of axon membrane for ion $i$, \\
				$J_{a,b}^{i}$: Ion $i$ flux across the interface form compartment $a$ to $b$, & $K_{k}$: Stiffness constant of the interface for compartment $k$, \\
				$J_{a,b}^{p,i}$: Active ATP based ion $i$ pump form compartment $a$ to $b$, & $\tau_{l}$: Tortuosity of compartment $l$, \\
				$J_{a,b}^{c,i}$: Passive source form compartment $a$ to $b$, & $D_{l}^{i}$: Diffusion coefficient of $i$ ion in compartment $l$, \\
				$I_{a,1}$: Max current of $\alpha_{1}-Na/k$ pump on $a$ membrane, & $T$: Temperature, \\
				$I_{a,2}$: Max current of $\alpha_{2}-Na/k$ pump on $a$ membrane, & $k_{B}$: Boltzmann constant, \\
				$A_{l}$: Negative charged protein density in compartment $l$, & $e$: The magnitude of the electron charge,\\
				$N_{A}$: Avogadro constant. &  
			\end{tabular}
		}
	\end{table}
	
	\begin{figure}
		\centering
		\includegraphics[width=0.45\linewidth]{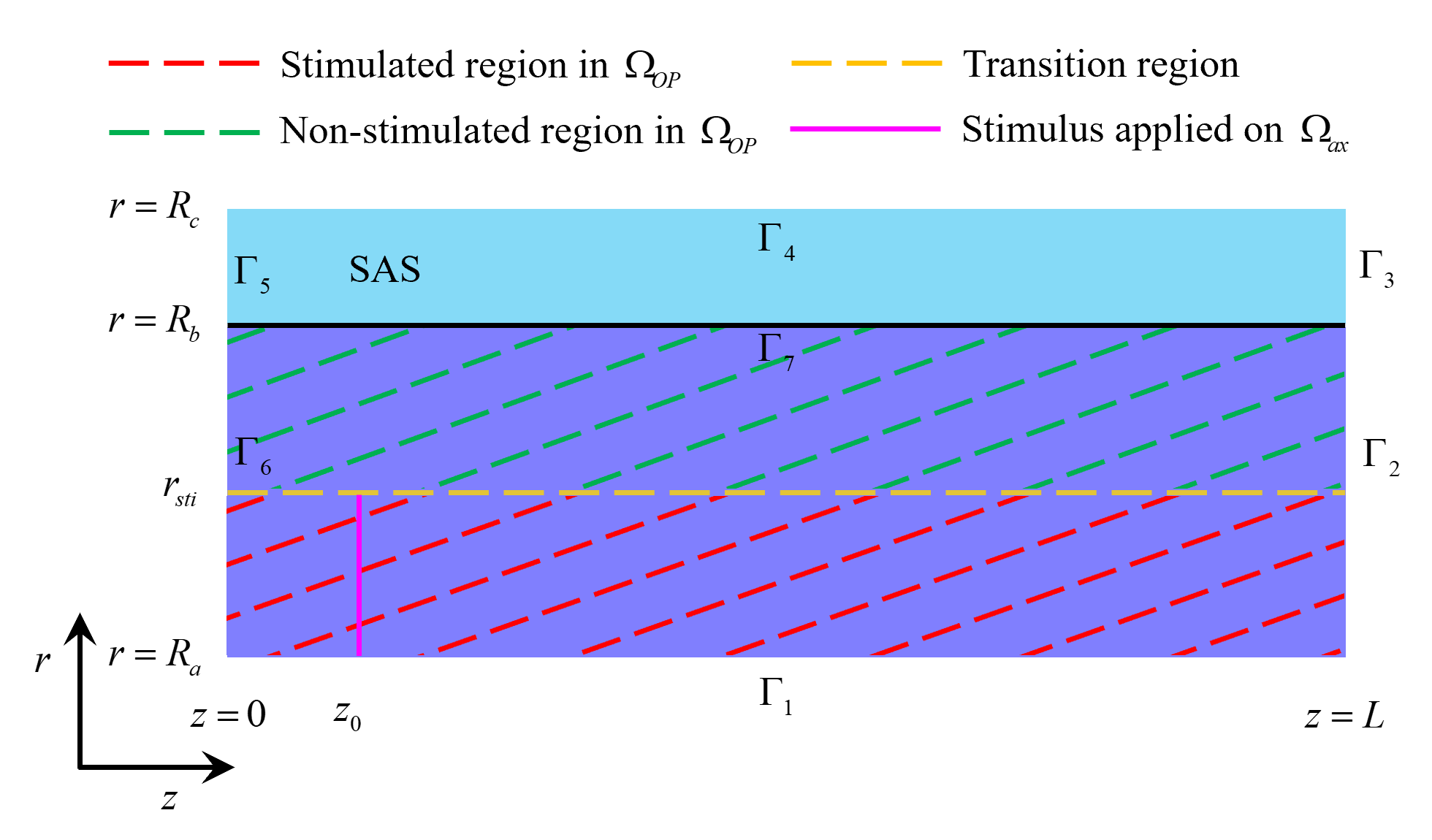}
		\includegraphics[width=0.45\linewidth]{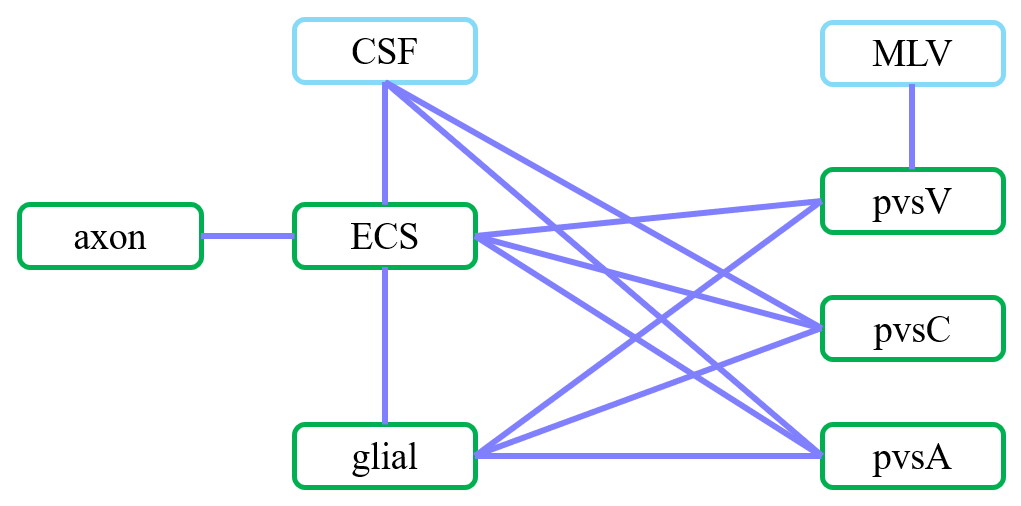}
		\caption{\label{fig:Op_structure_model} Top: The optic nerve $\Omega_{OP}$ consists of the axon compartment, glial compartment, extracellular space, and perivascular spaces. The subarachnoid space $\Omega_{SAS}$ only has cerebrospinal fluid.  The stimulus is applied to the axon membrane in the region $R_{a}<r<r_{sti}$ at location $z=z_{0}$. Bottom: The interaction between different regions. In the optic nerve $\Omega_{OP}$ region, the ECS exchanges fluid and ions with the axon, glial, the PVS-A, the PVS-V and the PVS-C; glial exchanges fluid and ion with the ECS, the PVS-A, the PVS-V and the PVS-C; CSF exchanges fluid and ions with the ECS, the PVS-A, and the PVS-C by across the pia mater in $\Gamma_{7}$.}
	\end{figure}
	
	The optic nerve is a complex structure consisting of multiple compartments responsible for ionic and fluid homeostasis. Figure~\ref{fig:Optic-nerve} illustrates the computational domain considered in this study. The domain, denoted as $\Omega$, consists of two main regions: the subarachnoid space (SAS), $\Omega_{\text{SAS}}$, and the optic nerve, $\Omega_{\text{OP}}$. Thus, we define:
	
	\begin{equation}
		\Omega = \Omega_{\text{OP}} \cup \Omega_{\text{SAS}}, \quad \Omega_{\text{OP}} \cap \Omega_{\text{SAS}} = \Gamma_7,
	\end{equation}
	
	\noindent where the  SAS region ($\Omega_{\text{SAS}}$) is filled with cerebrospinal fluid (CSF) and enclosed by the dura mater ($\Gamma_7$) and pia mater ($\Gamma_4$).
	
	The optic nerve microenvironment comprises multiple interconnected compartments, each playing a role in fluid and ion transport. Based on the six-domain microcirculation model introduced in \cite{2020ATridomain}, we define the optic nerve domain as:
	
	\begin{equation}
		\Omega_{\text{OP}} = \Omega_{\text{ax}} \cup \Omega_{\text{gl}} \cup \Omega_{\text{ex}} \cup \Omega_{\text{pa}} \cup \Omega_{\text{pv}} \cup \Omega_{\text{pc}},
	\end{equation}
\noindent  where: $\Omega_{\text{ax}}$ is  axon compartment; $\Omega_{\text{gl}}$ is  glial cell compartment; $\Omega_{\text{ex}}$ denotes extracellular space (ECS); $\Omega_{\text{pa}}$ denotes perivascular space surrounding arteries (PVS-A); $\Omega_{\text{pv}}$ denotes perivascular space surrounding veins (PVS-V); $\Omega_{\text{pc}}$ denotes perivascular space surrounding capillaries (PVS-C). 
	
	 {The boundaries of the computational domain are defined as follows: $\Gamma_1$ is the central retinal blood vessel wall; $\Gamma_2, \Gamma_3$ are distal ends of the optic nerve, connecting to the optic canal \cite{1984Thesheath}; $\Gamma_5$ denotes the dura mater connected to the sclera, assumed to be impermeable \cite{2009Ischemic} and $\Gamma_6$ denotes the left boundary near Lamina cribrosa, where the optic nerve head exits the eye through perforations \cite{2003Anatomic}.

	The model is derived from conservation laws governing ion and fluid transport across cellular membranes and extracellular compartments. For each domain $\Omega_{l}$, where $l = \text{ax}, \text{gl}, \text{ex}, \text{pa}, \text{pv}, \text{pc}$, we impose the general conservation equation:
	
	\begin{equation}
		\frac{\partial}{\partial t} (\eta_l f_l) + \nabla \cdot (\eta_l \mathbf{J}_l) + S = 0,
	\end{equation}
	
	\noindent where: $\eta_l$ is the  volume fraction of compartment $l$, $f_l$ is the concentration of a given solute, $\mathbf{J}_l$ is the flux within the compartment, and $S$ represents the source term due to trans-domain transport (e.g., active pumps, passive leak channels).
	
    We adopt the following assumptions to ensure computational feasibility and physiological relevance:
	
	\begin{itemize}
		\item \textbf{Axial symmetry:} The model assumes axial symmetry to reduce computational complexity while preserving essential transport dynamics. To clarify, the model is fundamentally three-dimensional. The axial symmetry assumed in our current implementation simplifies the numerical computation without altering the underlying physics. Importantly, the model formulation itself does not rely on symmetry assumptions and can be extended to fully three-dimensional, non-axisymmetric geometries to accommodate spatial heterogeneity or more anatomically realistic structures. 
		
		\item \textbf{Ion selection:} Only the three dominant ions involved in neural activity $\rm{Na^+}, \rm{K^+}$, and $\rm{Cl^-}$ are considered.
		
		\item \textbf{Charge neutrality:} Each compartment satisfies electroneutrality \cite{1999Transport}:
		
		\begin{subequations}\label{eq:charge_neutrality}
			\begin{align}
				& \eta_{\text{gl}} \sum_i z^i C_{\text{gl}}^i + z^{\text{gl}} \eta_{\text{gl}}^{\text{re}} A_{\text{gl}} = 0,\\
				& \eta_{\text{ax}} \sum_i z^i C_{\text{ax}}^i + z^{\text{ax}} \eta_{\text{ax}}^{\text{re}} A_{\text{ax}} = 0,\\
				& \sum_i z^i C_{l}^i = 0, \quad l = \text{ex}, \text{pa}, \text{pv}, \text{pc}, \text{csf}.
			\end{align}
		\end{subequations}
		
		Here, $A_{l}>0$ represents the protein density in axons and glial cells, which are permanently negatively charged but expressed as positive values for computational purposes. These proteins are assumed to be uniformly distributed at resting state with valences $z_{\text{ax}}^{-1}$ and $z_{\text{gl}}^{-1}$. The volume fractions $\eta_{\text{ax}}$ and $\eta_{\text{gl}}$ refer to their respective compartments, with $\eta_{\text{ax}}^{\text{re}}$ and $\eta_{\text{gl}}^{\text{re}}$ denoting their resting state values.
		
		\item \textbf{Anisotropy in axons, isotropy in other compartments:} 
		Axons are anisotropic, meaning ion and fluid transport occurs primarily along the axial direction due to their cylindrical structure and electrical isolation.
		- Other compartments (e.g., glial cells and ECS) are isotropic, allowing for both axial and radial diffusion and flow. Glial cells form a syncytium via connexins, facilitating intercellular ion and fluid exchange.
		
		\item \textbf{Compartmental interactions:} The interactions between different compartments, as illustrated in Figure~\ref{fig:Op_structure_model}, occur primarily through ion concentration gradients, electrical potentials, and trans-domain transport. There is no direct interaction between the axonal and glial compartments; instead, their communication is mediated through changes in the extracellular space (ECS) \cite{Sibille2015The}.
	\end{itemize}
	
	\subsection{Fluid Circulation}
	In this subsection, we present the fluid circulation model.  First, due to the conservation law, the volume fraction of each compartment $\eta_l, l= ax, gl,ex, pa,pv,pc$ satisfy
	
	\begin{subequations}
		\begin{align}
			& \dfrac{\partial \eta_{ax}}{\partial t}+\mathcal{M}_{ax,ex}U_{ax,ex}+\dfrac{\partial}{\partial z}(\eta_{ax}u_{ax}^{z})=0, \\
			\begin{split}
				& \dfrac{\partial \eta_{gl}}{\partial t}+\mathcal{M}_{gl,ex}U_{gl,ex}+\mathcal{M}_{gl,pa}U_{gl,pa}+\mathcal{M}_{gl,pv}U_{gl,pv}+\mathcal{M}_{gl,pc}U_{gl,pc}+\nabla\cdot(\eta_{gl}\mathbf{u}_{gl})=0, \\
			\end{split}\\
			\begin{split}
				& \dfrac{\partial \eta_{pa}}{\partial t}+\mathcal{M}_{pa,ex}U_{pa,ex}-\mathcal{M}_{gl,pa}U_{gl,pa}+\mathcal{M}_{pa,pc}U_{pa,pc}+\nabla\cdot(\eta_{pa}\mathbf{u}_{pa})=0,\\
			\end{split}\\
			\begin{split}
				& \dfrac{\partial \eta_{pv}}{\partial t}+\mathcal{M}_{pv,ex}U_{pv,ex}-\mathcal{M}_{gl,pv}U_{gl,pv}-\mathcal{M}_{pc,pv}U_{pc,pv}+\nabla\cdot(\eta_{pv}\mathbf{u}_{pv})=0,\\
			\end{split}\\
			\begin{split}
				& \dfrac{\partial \eta_{pc}}{\partial t}+\mathcal{M}_{pc,ex}U_{pc,ex}-\mathcal{M}_{gl,pc}U_{gl,pc}-\mathcal{M}_{pa,pc}U_{pa,pc}+\mathcal{M}_{pc,pv}U_{pc,pv}+\nabla\cdot(\eta_{pc}\mathbf{u}_{pc})=0,\\
			\end{split} \\
			\begin{split}
				& \dfrac{\partial}{\partial z}(\eta_{ax}u_{ax}^{z}) + \sum_{k=gl,ex,pa,pv,pc}\nabla\cdot(\eta_{k}\mathbf{u}_{k})=0,\\
			\end{split}\\
			& \eta_{ax}+\eta_{gl}+\eta_{ex}+\eta_{pa}+\eta_{pv}+\eta_{pc}=1,
		\end{align}
	\end{subequations}
	\noindent where $U_{l,k}$ is the fluid velocity across the membrane/interface between $l_{th}$ and $k_{th}$ compartments with  surface volume ratio $\mathcal{M}_{l,k}$ and $\mathbf{u}_l$ is the fluid velocity inside the $l_{th}$ compartment. 
	
	The trans-domain fluid flux is proportional to the intracellular/extracellular hydrostatic pressure and osmotic pressure differences, i.e., Starling's law on the membrane, while the fluid flow from PVS-A to PVS-C and from PVS-C to PVS-V are only proportional to the difference of hydrostatic pressure due to the direct connection.
	
	\begin{subequations}
		\begin{align}
			& U_{l,ex}=L_{l,ex}(p_{l}-p_{ex}-\gamma_{l,ex}RT(O_{l}-O_{ex})),~l=ax,gl,pa,pv,pc \\
			& U_{gl,l}=L_{gl,l}(p_{gl}-p_{l}-\gamma_{gl,l}RT(O_{gl}-O_{l})),~l=pa,pv,pc. 
		\end{align}
	\end{subequations}
	Here  $\gamma_{l,k}$  and $L_{l,k}$ are the reflection coefficient \cite{2017Quantitative} and hydraulic permeability of the membrane between $l_{th}$ and $k_{th}$ compartments, respectively. In this paper osmotic pressure is defined by $RTO_l$ \cite{2019ABidomain,2018Osmosis}  
	$$O_{l}=\sum_i C_{l}^i+A_{l}\dfrac{\eta_{l}^{re}}{\eta_{l}},  \quad  l=ax, gl,$$
	$$O_{l}=\sum_i C_{l}^i, \quad  l=ex, pa, pv, pc,$$
	where $A_{l}\dfrac{\eta_{l}^{re}}{\eta_{l}}>0$ is the density of the permanent negatively charged protein in glial cells and axons that varies with the volume (fraction) of the region,  $R$ is the molar gas constant, and $T$ is temperature.
	
	The hydrostatic pressure $p_l$ and the volume fraction $\eta_{l}$ are connected by the force balance on the membrane \cite{2018Osmosis,Mori2015A}. The membrane force is balanced by the hydrostatic pressure difference on both sides of the semipermeable membrane. Then the variation of volume fraction from the resting state is proportional to the variation of hydrostatic pressure difference from the resting state.
	
	\begin{equation}\label{eq:volumefraction}
		K_{l}(\eta_{l}-\eta_{l}^{re})=p_{l}-p_{ex}-(p_{l}^{re}-p_{ex}^{re})
	\end{equation}
	
	\noindent where $K_{l}$ is the stiffness constant and $\eta_{l}^{re}$ and $p_l^{re}$ are the resting state volume fraction and hydrostatic pressure of $l_{th}$ compartment with $l=ax,gl,pa,pv,pc$.
	
	The subarachnoid space region is modeled as a porous media filled with CSF. Therefore, the solution is incompressible in the $\Omega_{SAS}$, and we have 
	
	\begin{equation}
		\nabla\cdot\mathbf{u}_{csf}=0, \quad in \ \Omega_{SAS}.
	\end{equation}
	
	In the next section, we provide submodels of fluid velocities inside each compartment and the corresponding boundary conditions. 
	
	In this paper, we define a membrane boundary condition to describe the fluid or ion communication in the interface that connects the optic nerve and the retina or the optic canal or orbital. We assume this fluid or ion communication depends on the difference in pressure or ion concentration between the two sides of this connect interface $U_{int}=L(P_{in}-P_{out})$. Where $U_{int}$ is the interface velocity, $P_{in}$ and $P_{out}$ are hydrostatic pressure on both sides of the interface, respectively.
	
	\textbf{Fluid Velocity in the Glial Compartment.} The glial space is a connected space, where the intracellular fluid can flow from cell to cell through connexin proteins joining membranes of neighboring cells. The velocity of fluid in glial syncytium $\mathbf{u}_{gl}$ depends on the gradients of hydrostatic pressure and osmotic pressure \cite{2019ABidomain,2018Osmosis,Mori2015A,1985Steady,1979Electrical} as 
	
	\begin{subequations}
		\begin{align}
			u_{gl}^{r}& =-\dfrac{\kappa_{gl}\tau_{gl}}{\mu}\left(\dfrac{\partial p_{gl}}{\partial r}-\gamma_{gl}RT\dfrac{\partial O_{gl}}{\partial r}\right),\\
			u_{gl}^{\theta}& =-\dfrac{\kappa_{gl}\tau_{gl}}{\mu}\left(\dfrac{1}{r}\dfrac{\partial p_{gl}}{\partial \theta}-\gamma_{gl}RT\dfrac{1}{r}\dfrac{\partial O_{gl}}{\partial \theta}\right),\\
			u_{gl}^{z}& =-\dfrac{\kappa_{gl}\tau_{gl}}{\mu}\left(\dfrac{\partial p_{gl}}{\partial z}-\gamma_{gl}RT\dfrac{\partial O_{gl}}{\partial z}\right),
		\end{align}
	\end{subequations}
	where $\kappa_{gl}$ and $\tau_{gl}$  are the permeability and tortuosity of the glial compartment.

	For the boundary condition, on the left and right boundaries of domain $\Omega_{OP}$, the membrane boundary condition is used since they connect to intraocular and intracanalicular regions. On the top and bottom boundary, the no-flux boundary condition is used 
	
	\begin{equation}
		\begin{cases}
			\mathbf{u}_{gl}\cdot\hat{\mathbf{n}}_{r}=0, & \mbox{on} \ \Gamma_1\\
			\mathbf{u}_{gl}\cdot\hat{\mathbf{n}}_{z}=L_{gl,right}(p_{gl}-p_{gl,right}),& \mbox{on} \ \Gamma_2\\
			\mathbf{u}_{gl}\cdot\hat{\mathbf{n}}_{z}=L_{gl,left}(p_{gl}-p_{gl,left}), & \mbox{on} \ \Gamma_6\\
			\mathbf{u}_{gl}\cdot\hat{\mathbf{n}}_{r}=0, & \mbox{on} \ \Gamma_7\\
		\end{cases}
	\end{equation}
	where $\hat{\mathbf{n}}_r$ and $\hat{\mathbf{n}}_{z}$ are the outward normal vector of domain $\Omega_p$.
	
	\textbf{Fluid Velocity in the Axon Compartment.}  Since the axons are only connected in the longitudinal direction, the fluid velocity in the region of the axon is defined  as
	
	\begin{subequations}
		\begin{align}
			u_{ax}^{r}& =0,\\
			u_{ax}^{\theta}& =0,\\
			u_{ax}^{z}& =-\dfrac{\kappa_{ax}}{\mu}\dfrac{\partial p_{ax}}{\partial z},
		\end{align}
	\end{subequations}
	where $\kappa_{ax}$ is the permeability  of the axon compartment. 
	
	Similarly, membrane boundary conditions are used on the left and right boundaries 
	
	\begin{equation}
		\begin{cases}
			\mathbf{u}_{ax}\cdot\hat{\mathbf{n}}_{z}=L_{ax,right}(p_{ax}-p_{ax,right}), & \mbox{on} \ \Gamma_2\\
			\mathbf{u}_{ax}\cdot\hat{\mathbf{n}}_{z}=L_{ax,left}(p_{ax}-p_{ax,left}), & \mbox{on} \ \Gamma_6\\
		\end{cases}
	\end{equation}
	
	\textbf{Fluid Velocity in the Extracellular and Perivascular Spaces}  Since both the extracellular and perivascular spaces  are narrow, the  velocity is determined by the gradients of hydro-static pressure and electric potential \cite{2019ABidomain,2014Self,2012Development}, for $l=ex,pa,pv,pc$,
	\begin{subequations}
		\begin{align}
			u_{l}^{r}& =-\dfrac{\kappa_{l}\tau_{l}}{\mu}\dfrac{\partial p_{l}}{\partial r}-k_e^{l}\tau_{l}\dfrac{\partial \phi_{l}}{\partial r},\\
			u_{l}^{\theta}& =-\dfrac{\kappa_{l}\tau_{l}}{\mu}\dfrac{1}{r}\dfrac{\partial p_{l}}{\partial \theta}-k_e^{l}\tau_{l}\frac{1}{r}\dfrac{\partial \phi_{l}}{\partial \theta},\\
			u_{l}^{z}& =-\dfrac{\kappa_{l}\tau_{l}}{\mu}\dfrac{\partial p_{l}}{\partial z}-k_e^{l}\tau_{l}\dfrac{\partial \phi_{l}}{\partial z},
		\end{align}
	\end{subequations}
	where $\phi_{l}$ is the electric potential, $\tau_{l}$ is the tortuosity \cite{1983Effect,2001Diffusion,1995Extracellular}, $k_e^l$ describes the effect of electro-osmotic flow \cite{1977Electrical,2014Self,2012Development,1985Electro}, $\kappa_{l}$ is the permeability. 
	
	For the boundary conditions, due to the connections to intraocular region and intracanalicular region on the left and right boundaries, respectively,  membrane boundary conditions are used for extracellular and perivascular spaces A \& V. Non-flux boundary condition is used for PVS-C on $\Gamma_2$ and $\Gamma_6$. On $\Gamma_1$, due to the central blood vessels, Dirichlet boundary conditions on pressure are used for  PVS-A and PVS-V.  For the ECS and PVS-C, the zero penetration velocity is used. On the pia mater $\Gamma_7$, the CSF could directly communicate with the perivascular spaces A due to the hydrostatic pressure difference; membrane boundary condition is used for the perivascular spaces V due to the connections to PVS-V and the MLV. However, for ECS and PVS-C, the CSF leaks into these two compartments through membranes. Then the velocity is fixed as the trans-domain velocity which depends on hydrostatic and osmotic pressure difference.  In summary, the conditions are listed as follows
	
	\begin{equation}
		\begin{cases}
			p_{pa}=p_{PA},~ p_{pv}=p_{PV}, ~\mathbf{u}_{pc}\cdot\hat{\mathbf{n}}_{r}=0, ~\mathbf{u}_{ex}\cdot\hat{\mathbf{n}}_{r}=0,& \mbox{on} \ \Gamma_1\\
			\mathbf{u}_{pa}\cdot\hat{\mathbf{n}}_{z}=L_{pa,right}(p_{pa}-p_{pa,right}), \mathbf{u}_{pv}\cdot\hat{\mathbf{n}}_{z}=L_{pv,right}(p_{pv}-p_{pv,right}),   & \mbox{on}  \ \Gamma_2\\
			\mathbf{u}_{pc}\cdot\hat{\mathbf{n}}_{z}=0,\mathbf{u}_{ex}\cdot\hat{\mathbf{n}}_{z}=L_{ex,right}(p_{ex}-p_{pv,right}), & \mbox{on}  \ \Gamma_2\\
			\mathbf{u}_{pa}\cdot\hat{\mathbf{n}}_{z}=L_{pa,left}(p_{pa}-p_{pa,left}),\mathbf{u}_{pv}\cdot\hat{\mathbf{n}}_{z}=L_{pv,left}(p_{pv}-p_{IOP}), & \mbox{on}  \ \Gamma_6\\
			\mathbf{u}_{pc}\cdot\hat{\mathbf{n}}_{z}=0, \mathbf{u}_{ex}\cdot\hat{\mathbf{n}}_{z}=L_{ex,left}(p_{ex}-p_{IOP}), & \mbox{on}  \ \Gamma_6\\
			\mathbf{u}_{pa}\cdot\hat{\mathbf{n}}_{r}=L_{pia,pa}(p_{pa}-p_{csf})), 	\mathbf{u}_{pv}\cdot\hat{\mathbf{n}}_{r}=L_{mlv,pv}(p_{pv}-p_{mlv})), & \mbox{on}  \ \Gamma_7\\
			\mathbf{u}_{pc}\cdot\hat{\mathbf{n}}_{r}=L_{pia,pc}(p_{pc}-p_{csf}-\gamma_{pia}RT(O_{pc}-O_{csf})), &\mbox{on}  \ \Gamma_7\\
			\mathbf{u}_{ex}\cdot\hat{\mathbf{n}}_{r}=L_{pia,ex}(p_{ex}-p_{csf}-\gamma_{pia}RT(O_{ex}-O_{csf})),  &\mbox{on}  \ \Gamma_7.
		\end{cases}
	\end{equation}
	
	\noindent where $p_{IOP}$ is the intraocular pressure (IOP), $\gamma_{pia}$ is the  the reflection coefficient of pia mater.
	
	\textbf{Fluid Velocity in the SAS Region.} The cerebrospinal fluid velocity in the SAS region is determined by the gradients of hydro-static pressure and electric potential
	
	\begin{subequations}
		\begin{align}
			u_{csf}^{r}& =-\dfrac{\kappa_{csf}\tau_{csf}}{\mu}\dfrac{\partial p_{csf}}{\partial r}-k_e^{csf}\tau_{csf}\dfrac{\partial \phi_{csf}}{\partial r},\\
			u_{csf}^{\theta}& =-\dfrac{\kappa_{csf}\tau_{csf}}{\mu}\dfrac{1}{r}\dfrac{\partial p_{csf}}{\partial \theta}-k_e^{csf}\tau_{csf}\frac{1}{r}\dfrac{\partial \phi_{csf}}{\partial \theta},\\
			u_{csf}^{z}& =-\dfrac{\kappa_{csf}\tau_{csf}}{\mu}\dfrac{\partial p_{csf}}{\partial z}-k_e^{csf}\tau_{csf}\dfrac{\partial \phi_{csf}}{\partial z},
		\end{align}
	\end{subequations}
	
	The fluid flow across the semi-permeable membrane $\Gamma_4$ is produced by the lymphatic drainage on the dura membrane, which depends on the difference between cerebrospinal fluid pressure and orbital pressure (OBP). At boundary $\Gamma_3$, we assume the hydrostatic pressure is equal to the cerebrospinal fluid pressure. At boundary $\Gamma_5$,  a non-permeable boundary is used.  On the pia membrane $\Gamma_7$,  the total CSF trans-domain velocity is determined by the conservation law, 
	
	\begin{equation}
		\begin{cases}
			p_{csf}=p_{CSF}, & \mbox{on} \ \Gamma_3\\
			\mathbf{u}_{csf}\cdot\hat{\mathbf{n}}_{r}=L_{dr}(p_{csf}-p_{OBP}), & \mbox{on} \ \Gamma_4\\
			\mathbf{u}_{csf}\cdot\hat{\mathbf{n}}_{z}=0, & \mbox{on} \ \Gamma_5\\
			\mathbf{u}_{csf}\cdot\hat{\mathbf{n}}_{r}=L_{pia,ex}(p_{ex}-p_{csf}-\gamma_{pia}RT(O_{ex}-O_{csf}))\\
			\ \ \ \ \ \ \ \ \ \ \ \ +L_{pia,pa}(p_{pa}-p_{csf}) \\
			\ \ \ \ \ \ \ \ \ \ \ \ +L_{pia,pc}(p_{pc}-p_{csf}-\gamma_{pia}RT(O_{pc}-O_{csf})), & \mbox{on} \ \Gamma_7\\
		\end{cases}
	\end{equation}
	
	\noindent where $p_{CSF}$ is the cerebrospinal fluid pressure \cite{2009Intracellular} in boundary $\Gamma_3$.
	
	\subsection{Ion Transport}
	
	The conservation of ion species implies the following system of partial differential equations to describe the dynamics of ions in each region, for $i=\rm{Na^+},~\rm{K^+},~\rm{Cl^-}$ in domain $\Omega_{OP}$
	
	\begin{subequations}\label{eq:Iontransport}
		\begin{align}
			& \dfrac{\partial }{\partial t}(\eta_{ax}C_{ax}^{i})+\mathcal{M}_{ax,ex}J_{ax,ex}^{i}+\dfrac{\partial}{\partial z}(\eta_{ax}j_{ax,z}^{i})=0,\\
			& \dfrac{\partial }{\partial t}(\eta_{gl}C_{gl}^{i})+\mathcal{M}_{gl,ex}J_{gl,ex}^{i}+\mathcal{M}_{gl,pa}J_{gl,pa}^{i}+\mathcal{M}_{gl,pv}J_{gl,pv}^{i}+\mathcal{M}_{gl,pc}J_{gl,pc}^{i}+\nabla\cdot(\eta_{gl}\mathbf{j}_{gl}^{i})=0,\\
			& \dfrac{\partial }{\partial t}(\eta_{pa}C_{pa}^{i})+\mathcal{M}_{pa,ex}J_{pa,ex}^{i}-\mathcal{M}_{gl,pa}J_{gl,pa}^{m,i}+\mathcal{M}_{pa,pc}J_{pa,pc}^{i}+\nabla\cdot(\eta_{pa}\mathbf{j}_{pa}^{i})=0,\\
			& \dfrac{\partial }{\partial t}(\eta_{pv}C_{pv}^{i})+\mathcal{M}_{pv,ex}J_{pv,ex}^{i}-\mathcal{M}_{gl,pv}J_{gl,pv}^{m,i}-\mathcal{M}_{pc,pv}J_{pc,pv}^{i}+\nabla\cdot(\eta_{pv}\mathbf{j}_{pv}^{i})=0,\\
			& \dfrac{\partial }{\partial t}(\eta_{pc}C_{pc}^{i})+\mathcal{M}_{pc,ex}J_{pc,ex}^{i}-\mathcal{M}_{gl,pc}J_{gl,pc}^{i}-\mathcal{M}_{pa,pc}J_{pa,pc}^{i}+\mathcal{M}_{pc,pv}J_{pc,pv}^{i}+\nabla\cdot(\eta_{pc}\mathbf{j}_{pc}^{i})=0,\\
			& \dfrac{\partial }{\partial t}(\eta_{ex}C_{ex}^{i})-\sum_{k=ax,gl,pa,pv,pc}\mathcal{M}_{k,ex}J_{k,ex}^{i}+\nabla\cdot(\eta_{ex}\mathbf{j}_{ex}^{i})=0,
		\end{align}
	\end{subequations}
	and in the $\Omega_{SAS}$ region,
	\begin{equation}
		\dfrac{\partial C_{csf}^{i}}{\partial t}+\nabla\cdot(\mathbf{j}_{csf}^{i})=0.
	\end{equation}
	
	\paragraph{trans-domain Ion Flux}
	The trans-domain ion flux $J_{l,k}^{i}$ ( $l,k=ax,ex;gl,ex;gl,pa;gl,pv;gl,pc;$) consists of an active ion pump source $J_{l,k}^{p,i}$ and passive ion channel source $J_{l,k}^{c,i}$,
	
	$$J_{l,k}^{i}=J_{l,k}^{p,i}+J_{l,k}^{c,i};\ \  i=\rm{Na^+},~\rm{K^+},~\rm{Cl^-}.$$
	
	Due to the gaps between astrocytes endfeet \cite{ray2021quantitative}, 	the trans-domain ion flux $J_{l,k}^{i}$ between perivascular spaces and ECS $(l,k=pa,ex;pv,ex;pc,ex)$ consists of direct transportation with fluid $C_{up,wind}^{i}U_{l,k}$ and passive ion channel source $J_{l,k}^{c,i}$,
	
	$$J_{l,k}^{i}=C_{up,wind}^{i}U_{l,k}+J_{l,k}^{c,i};\ \  i=\rm{Na^+},~\rm{K^+},~\rm{Cl^-}.$$
	
	Because of the direct connection, the communication ion flux $J_{l,k}^{i}$   between perivascular spaces only depends on transportation with fluid $C_{up,wind}^{i}U_{l,k},$
	
	$$J_{l,k}^{i}=C_{up,wind}^{i}U_{l,k};\ \  i=\rm{Na^+},~\rm{K^+},~\rm{Cl^-}.$$
	
	On the glial cell membranes, $J_{gl,k}^{c,i}$ is defined as
	
	\begin{equation}\label{Jtransmem_gl}
		J_{gl,k}^{c,i}=\dfrac{g_{gl}^{i}}{z^{i}e}(\phi_{gl}-\phi_{k}-E_{gl,k}^{i}),\ i=\rm{Na^+,K^+,Cl^-}, \ k=ex,pa,pv,pc,
	\end{equation}
	
	\noindent where $E_{gl,k}^{i}$ is the Nernst potential that describes the gradient of chemical potential in electrical units
	$$E_{gl,k}^{i}=\dfrac{k_BT}{ez^{i}}log\left(\dfrac{C_{k}^{i}}{C_{gl}^{i}}\right), k=ex,pa,pv,pc$$  and the conductance $g_{gl}^{i}$ for the $i$th ion species on the glial membrane is a fixed constant, independent of voltage and time. 
	
	On the axon membrane, $J_{ax,ex}^{c,i}$ is definded as 
	
	\begin{equation}
		J_{ax,ex}^{c,i}=\dfrac{g_{ax}^{i}}{z^{i}e}(\phi_{ax}-\phi_{ex}-E_{ax}^{i}),\ i=\rm{Na^+,K^+,Cl^-},
	\end{equation}
	
	\noindent where the conductances of $\rm{Na^+}$ and $\rm{K^+}$ are modeled using  the Hodgkin-Huxley model \cite{1960Thresholds,2017Mathematics}
	
	$$g_{ax}^{Na}=\bar{g}^{Na}m^{3}h+g_{leak}^{Na},\ g_{ax}^{K}=\bar{g}^{K}n^{4}+g_{leak}^{K},\ g_{ax}^{Cl}=g_{leak}^{Cl}.$$
	
	\begin{equation}
			\dfrac{dn}{dt}=\alpha_{n}(1-n)-\beta_{n}n,~~
			\dfrac{dm}{dt}=\alpha_{m}(1-m)-\beta_{m}m,~~
			\dfrac{dh}{dt}=\alpha_{h}(1-h)-\beta_{h}h,
	\end{equation}
	
	\noindent where $n$ is the open probability of $K^{+}$ channel, $m$ is the open probability of the $Na^{+}$ activation gate, and $h$ is the open probability of the $Na^{+}$ inactivation gate. $\alpha_i$ and $\beta_i$ for $i=n, m, h$ are active and inactive rates of different gate. 
	
	For the active ion pump source $J_{l,k}^{p,i}$, the only pump we consider is the Na/K active transporter. We are more than aware that other active transport systems can and likely do move ions and fluid in this system. They will be included as experimental information becomes available. In the case of the Na/K pump $J_{l,k}^{p,i},\ (l,k=ax,ex;gl,pa;gl,pv;gl,pc)$,  the strength of the pump depends on the concentration in the intracellular and extracellular spaces \cite{1960Thresholds,2000Isoform}, such that 
	
	\begin{equation}
		J_{l,k}^{p,Na}=\dfrac{3I_{l}}{e},\ J_{l,k}^{p,K}=-\dfrac{2I_{l}}{e},\ J_{l,k}^{p,Cl}=0,\ l=ax,gl,
	\end{equation}
	
	\noindent where 
	
	$$I_{l}=I_{l,1}\left(\dfrac{C_{l}^{Na}}{C_{l}^{Na}+K_{Na1}}\right)^{3}\left(\dfrac{C_{ex}^{K}}{C_{ex}^{K}+K_{K1}}\right)^{2}+I_{l,2}\left(\dfrac{C_{l}^{Na}}{C_{l}^{Na}+K_{Na2}}\right)^{3}\left(\dfrac{C_{ex}^{K}}{C_{ex}^{K}+K_{K2}}\right)^{2},$$
	
	\noindent $I_{l,1}$ and $I_{l,2}$ are related to $\alpha_{1}-$ and $\alpha_{2}-$ isoform of Na/K pump. 
	
	\paragraph{Ion Flux inside Compartment }	The definitions of ion flux in each domain are as follows, for $i=\rm{Na^+,~K^+,~Cl^-},$
	
	\begin{subequations}
		\begin{align}
			& \mathbf{j}_{l}^{i}=C_{l}^{i}\mathbf{u}_{l}-D_{l}^{i}\tau_{l}\left(\nabla C_{l}^{i}+\dfrac{z^{i}e}{k_{B}T}C_{l}^{i}\nabla\phi_{l}\right),\ l=gl,ex,pa,pv,pc, \\
			& j_{ax,z}^{i}=C_{ax}^{i}u_{ax}^{z}-D_{ax}^{i}\left(\dfrac{\partial C_{ax}^{i}}{\partial z}+\dfrac{z^{i}e}{k_{B}T}C_{ax}^{i}\dfrac{\partial\phi_{ax}}{\partial z}\right).\ \ \ \ \ \ \ \ \ \ \ \ \ \ \ \ \ \ \ \ \ \ \ \ \ \ \ \ \ \ \        
		\end{align}
	\end{subequations}
	\paragraph{Boundary Conditions} 	
	For the axon and glial compartment boundary condition, we use the membrane boundary conditions at location $\Gamma_2\cup\Gamma_6$. The homogeneous Neumann boundary condition on the $\Gamma_1$ and a non-flux boundary condition is used on the pia mater $\Gamma_7$ for the glial compartment.
	
	\begin{equation}
		\begin{cases}
			\nabla C_{gl}^{i}\cdot\hat{\mathbf{n}}_{r}=0, & \mbox{on} \ \Gamma_1,\\
			\nabla C_{ax}^{i}\cdot\hat{\mathbf{n}}_{z}=\lambda_{ax,left}(C_{ax}^{i}-C_{ax}^{i,re}),\ \nabla C_{gl}^{i}\cdot\hat{\mathbf{n}}_{z}=\lambda_{gl,left}(C_{gl}^{i}-C_{gl}^{i,re}), & \mbox{on} \ \Gamma_2,\\
			\nabla C_{ax}^{i}\cdot\hat{\mathbf{n}}_{z}=\lambda_{ax,right}(C_{ax}^{i}-C_{ax}^{i,re}),\ \nabla C_{gl}^{i}\cdot\hat{\mathbf{n}}_{z}=\lambda_{gl,right}(C_{gl}^{i}-C_{gl}^{i,re}), & \mbox{on} \ \Gamma_6,\\
			\mathbf{j}_{gl}^{i}\cdot\hat{\mathbf{n}}_{r}=0, & \mbox{on} \ \Gamma_7,\\
		\end{cases}
	\end{equation}
	where $\lambda_{ax(gl), left(right)}$ is the ion communication rate on the left (right) boundary of the axon (glial) compartment.

	We use the Dirichlet boundary conditions for the PVS-A/V boundary condition at location $\Gamma_1$. For the PVS-C and the ECS boundary condition, we use the homogeneous Neumann boundary condition at location $\Gamma_1$. The membrane boundary conditions at locations $\Gamma_2\cup\Gamma_6$ are used for the PVS-A/V and the ECS. And the homogeneous Neumann boundary condition on $\Gamma_2\cup\Gamma_6$ is used for the PVS-C. The flux across the pia mater is assumed continuous and Ohm's law \cite{2019ABidomain} and the additional pathway for diffusion, electric drift as well as convection for ions is used at location $\Gamma_7$ for the PVS-A/C and the ECS.

	\begin{equation}
		\begin{cases}
			C_{pa}^{i}=C_{pa}^{i,re},\ C_{pv}^{i}=C_{pv}^{i,re},\ \nabla C_{pc}^{i}\cdot\hat{\mathbf{n}}_{r}=0,\ \nabla C_{ex}^{i}\cdot\hat{\mathbf{n}}_{r}=0, & \mbox{on} \ \Gamma_1,\\
			\nabla C_{pa}^{i}\cdot\hat{\mathbf{n}}_{z}=\lambda_{pa,left}(C_{pa}^{i}-C_{pa}^{i,re}),\ \nabla C_{pv}^{i}\cdot\hat{\mathbf{n}}_{z}=\lambda_{pv,left}(C_{pv}^{i}-C_{pc}^{i,re}), & \mbox{on} \ \Gamma_2,\\
			\nabla C_{pc}^{i}\cdot\hat{\mathbf{n}}_{z}=0,\ \nabla C_{ex}^{i}\cdot\hat{\mathbf{n}}_{z}=\lambda_{ex,left}(C_{ex}^{i}-C_{ex}^{i,re}), & \mbox{on} \ \Gamma_2,\\
			\nabla C_{pa}^{i}\cdot\hat{\mathbf{n}}_{z}=\lambda_{pa,right}(C_{pa}^{i}-C_{pa}^{i,re}),\ \nabla C_{pv}^{i}\cdot\hat{\mathbf{n}}_{z}=\lambda_{pv,right}(C_{pv}^{i}-C_{pc}^{i,re}), & \mbox{on} \ \Gamma_6,\\
			\nabla C_{pc}^{i}\cdot\hat{\mathbf{n}}_{z}=0,\ \nabla C_{ex}^{i}\cdot\hat{\mathbf{n}}_{z}=\lambda_{ex,right}(C_{ex}^{i}-C_{ex}^{i,re}), & \mbox{on} \ \Gamma_6,\\
			\nabla C_{pv}^{i}\cdot\hat{\mathbf{n}}_{r}=0, \ & \mbox{on} \ \Gamma_{7},\\
			\mathbf{j}_{pa}^{i}\cdot\hat{\mathbf{n}}_{r}=\dfrac{G_{pia}^{i}}{z^{i}e}\left(\phi_{pa}-\phi_{csf}-\dfrac{k_BT}{ez^{i}}log\left(\dfrac{C_{csf}^{i}}{C_{pa}^{i}}\right)\right)+C_{pa}^{i}u_{pa,csf}, & \mbox{on} \ \Gamma_7,\\
			\mathbf{j}_{pc}^{i}\cdot\hat{\mathbf{n}}_{r}=\dfrac{G_{pia}^{i}}{z^{i}e}\left(\phi_{pc}-\phi_{csf}-\dfrac{k_BT}{ez^{i}}log\left(\dfrac{C_{csf}^{i}}{C_{pc}^{i}}\right)\right), & \mbox{on} \ \Gamma_7,\\
			\mathbf{j}_{ex}^{i}\cdot\hat{\mathbf{n}}_{r}=\dfrac{G_{pia}^{i}}{z^{i}e}\left(\phi_{ex}-\phi_{csf}-\dfrac{k_BT}{ez^{i}}log\left(\dfrac{C_{csf}^{i}}{C_{ex}^{i}}\right)\right), & \mbox{on} \ \Gamma_7.\\
		\end{cases}
	\end{equation}
	
	For the cerebrospinal fluid boundary condition, similar boundary conditions are imposed except on $\Gamma_5$, where a non-permeable boundary condition is used.
	
	\begin{equation}
		\begin{cases}
			C_{csf}^{i}=C_{csf}^{i,re}, & \mbox{on} \ \Gamma_3,\\
			\nabla C_{csf}^{i}\cdot\hat{\mathbf{n}}_{r}=0, & \mbox{on} \ \Gamma_4,\\
			\mathbf{j}_{csf}^{i}\cdot\hat{\mathbf{n}}_{z}=0, & \mbox{on} \ \Gamma_5,\\
			\mathbf{j}_{csf}^{i}\cdot\hat{\mathbf{n}}_{r}= \sum\limits_{l=pa,pc,ex} \mathbf{j}_{l}^{i}\cdot\hat{\mathbf{n}}_{r}, & \mbox{on} \ \Gamma_7.
		\end{cases}
	\end{equation}
	
	\subsection{Electric Potential}
	
	By multiplying equations \ref{eq:Iontransport} with $z^{i}e$ respectively, summing up, and using charge neutrality equation \ref{eq:charge_neutrality} and ion flux equation, we have the following system for the electric potential in $ax,gl,ex,pa,pv,pc$
	
	\begin{subequations}
		\begin{align}
			& \sum_{i}z^{i}e\left(\mathcal{M}_{ax,ex}J_{ax,ex}^{i}+\dfrac{\partial}{\partial z}(\eta_{ax}j_{ax,z}^{i})\right)=0,\\
			\begin{split}
				& \sum_{k=ex,pa,pv,pc}\left(\sum_{i}z^{i}e\left(\mathcal{M}_{gl,k}J_{gl,k}^{i}\right)\right)+\sum_{i}z^{i}e\left(\nabla\cdot(\eta_{gl}\mathbf{j}_{gl}^{i})\right)=0,   \\
			\end{split}\\
			\begin{split}
				& \sum_{i}z^{i}e\left(\mathcal{M}_{pa,ex}J_{pa,ex}^{i}-\mathcal{M}_{gl,pa}J_{gl,pa}^{i}+\mathcal{M}_{pa,pc}J_{pa,pc}^{i}+\nabla\cdot(\eta_{pa}\mathbf{j}_{pa}^{i})\right)=0,\\
			\end{split}\\
			\begin{split}
				& \sum_{i}z^{i}e\left(\mathcal{M}_{pv,ex}J_{pv,ex}^{i}-\mathcal{M}_{gl,pv}J_{gl,pv}^{i}-\mathcal{M}_{pc,pv}J_{pc,pv}^{i}+\nabla\cdot(\eta_{pv}\mathbf{j}_{pv}^{i})\right)=0,\\
			\end{split}\\
			\begin{split}
				& \sum_{i}z^{i}e\left(\mathcal{M}_{pc,ex}J_{pc,ex}^{i}-\mathcal{M}_{gl,pc}J_{gl,pc}^{i}-\mathcal{M}_{pa,pc}J_{pa,pc}^{i}+\mathcal{M}_{pc,pv}J_{pc,pv}^{i}+\nabla\cdot(\eta_{pc}\mathbf{j}_{pc}^{i})\right)=0,\
			\end{split}\\
			\begin{split}
				& \sum_{k=ax,gl,pa,pv,pc}\left(-\sum_{i}z^{i}e\left(\mathcal{M}_{k,ex}J_{k,ex}^{i}\right)\right)+\sum_{i}z^{i}e\left(\nabla\cdot(\eta_{ex}\mathbf{j}_{ex}^{i})\right)=0,\\
			\end{split}
		\end{align}
	\end{subequations}
	
	\noindent which describe the spatial distributions of electric potentials in six compartments.
	
	In the subarachnoid space $\Omega_{SAS}$,  the governing equation for cerebrospinal fluid electric potential reduces to
	
	\begin{equation}
		\sum_{i}z^{i}e\left(\nabla\cdot\mathbf{j}_{csf}^{i}\right)=0.
	\end{equation}
	
	The boundary conditions for electric fields $\phi_{ax},\phi_{gl},\phi_{pa},\phi_{pv},\phi_{pc},\phi_{ex},\phi_{csf}$ are given below.
	
	\begin{equation}
		\begin{cases}
			\nabla\phi_{l}\cdot\hat{\mathbf{n}}_{r}=0,\ l=gl,pa,pv,pc,ex, & \mbox{on} \ \Gamma_1,\\
			\nabla\phi_{l}\cdot\hat{\mathbf{n}}_{z}=0,\ l=ax,gl,pa,pv,pc,ex, & \mbox{on} \ \Gamma_2\cup\Gamma_6,\\
			\nabla\phi_{csf}\cdot\hat{\mathbf{n}}_{r}=0, & \mbox{on} \ \Gamma_3\cup\Gamma_5,\\
			\nabla\phi_{csf}\cdot\hat{\mathbf{n}}_{z}=0, & \mbox{on} \ \Gamma_4,\\
			\nabla\phi_{l}\cdot\hat{\mathbf{n}}_{r}=0,\ l=gl,pv, & \mbox{on} \ \Gamma_7,\\
			\sum_{i}z^{i}e\mathbf{j}_{l}^{i}\cdot\hat{\mathbf{n}}_{r}=\sum_{i}G_{pia}^{i}\left(\phi_{l}-\phi_{csf}-\dfrac{k_BT}{ez^{i}}log\left(\dfrac{C_{csf}^{i}}{C_{l}^{i}}\right)\right),\ l=pa,pc,ex, & \mbox{on} \ \Gamma_7,\\
			\sum_{i}z^{i}e\mathbf{j}_{csf}^{i}\cdot\hat{\mathbf{n}}_{r}= \sum\limits_{l=pa,pc,ex}\sum_{i}z^{i}e\mathbf{j}_{l}^{i}\cdot\hat{\mathbf{n}}_{r}& \mbox{on} \ \Gamma_7.\\
		\end{cases}
	\end{equation}
	
	\section{Fluid Circulation in the Optic Nerve}
	
	Fluid homeostasis in the central nervous system (CNS) is maintained by the coordinated movement of four primary fluid types: intracellular fluid (ICF) (60–68\%), interstitial/extracellular fluid (ISF/ECS) (12–20\%), blood (10\%), and cerebrospinal fluid (CSF) (10\%) \cite{2015TheGlymphatic}. This section examines fluid microcirculation in the optic nerve, both under resting conditions and in response to neuronal stimuli.

This study primarily focuses on fluid dynamics. Results related to ionic transport are included only where they directly influence osmotic pressure, electric potential gradients, or fluid flow mechanisms. A detailed analysis of ionic dynamics will be presented in a separate study.
	
	The model is solved using the Finite Volume Method on a uniform mesh with axial symmetry, with equal discretization in the radial and axial directions, i.e., $N_r=N_z=N=20$ and  the time step is fixed at $\delta t=10^{-1}$ in dimensionless units. As a convergence criterion, the simulation is considered to reach steady state when the maximum variation of all variables between two successive time steps falls below $10^{-8}$. To further ensure numerical stability, we adopt conservative flux formulations and implicit time stepping. All variables are monitored to remain within physiologically meaningful ranges throughout the simulation. The computational model was developed and executed in MATLAB. The resting-state equilibrium was determined iteratively by setting a fixed volume fraction for each compartment \cite{1966Physiological}:
	
	\[
	\eta_{ax}^{re}=0.4, \quad \eta_{gl}^{re}=0.4, \quad \eta_{ex}^{re}=0.1, \quad \eta_{pa}^{re}=0.024, \quad \eta_{pv}^{re}=0.0639, \quad \eta_{pc}^{re}=0.0121.
	\]
	
	These equilibrium values serve as the initial conditions for subsequent simulations.
	
	\subsection{Resting-State Fluid Circulation}
	
	\begin{figure}
		\centering
		\includegraphics[width=0.6\linewidth]{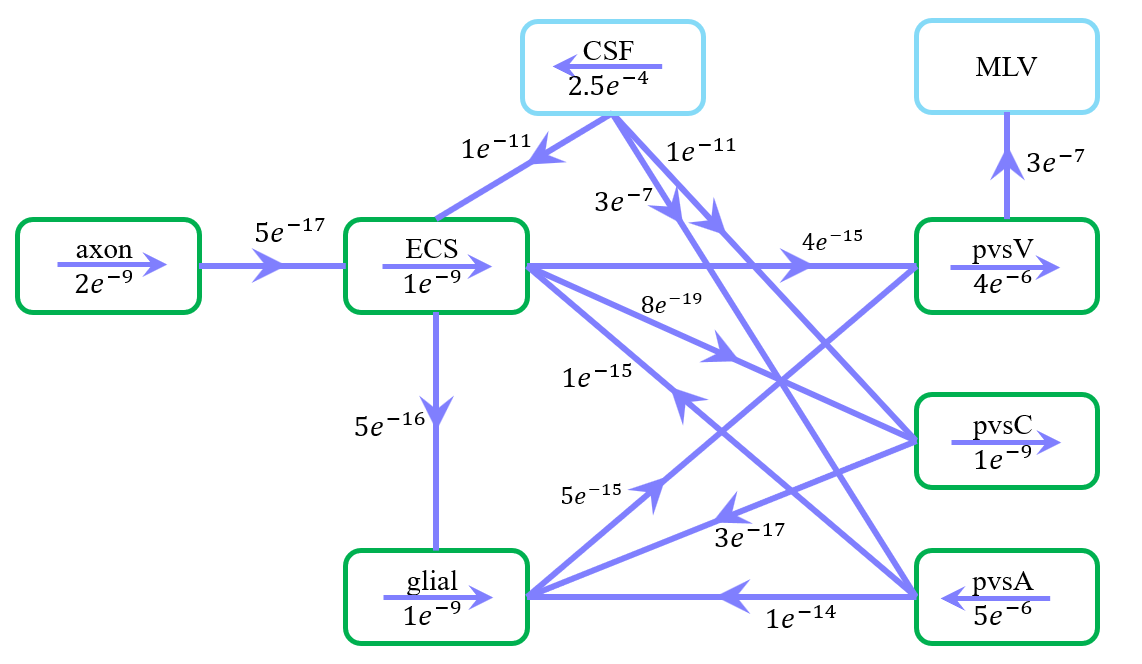}
		\caption{\label{fig:Velocitydirectionresting} Schematic of fluid flow within compartments and trans-domain fluid exchange in the resting state. Arrows indicate intra- and inter-compartmental flow directions, while numerical values represent spatially averaged velocity magnitudes ($\rm{m/s}$).}
	\end{figure}
	
	Figure~\ref{fig:Velocitydirectionresting} illustrates the velocity distribution across compartments in the resting state. Green arrows indicate flow direction, while numerical values represent average velocity magnitudes ($\rm{m/s}$).
	
	Under a pressure gradient of $0.0083~\rm{mmHg/mm}$, CSF enters the subarachnoid space (SAS) from the intracranial region, with an average velocity of $250 ~\rm{\mu m/s}$ in the z-direction, consistent with previous studies \cite{2021Lymphatics}. CSF is then transported into the extracellular space (ECS) and perivascular spaces (PVS-A and PVS-C) through astrocytic endfoot aquaporin-4 (AQP4) channels and paracellular gaps, subsequently flowing through the optic nerve and glial compartments. Eventually, the fluid converges in the PVS-V(PVS-V) and drains into the cervical lymphatic system \cite{uddin2022ocular}.
	
	Fluid flow within perivascular spaces is governed by pressure gradients and compartmental interactions:
	\begin{itemize}
		\item Flow in PVS-A aligns with the central retinal artery, while flow in PVS-V follows the central retinal vein.
		\item Fluid in PVS-C dynamically adjusts through exchange with adjacent compartments, primarily influenced by PVS-V due to the pressure gradient at boundary $\Gamma_1$.
		\item CSF exits the optic nerve at boundary $\Gamma_2$, with additional clearance via the cervical lymphatic system at $\Gamma_7$.
	\end{itemize}

	The spatially averaged velocity in PVS-A under a $0.007~\rm{mmHg/mm}$ pressure gradient is $5~\rm{\mu m/s}$, directed from the intracanalicular space to the intraorbital region. Conversely, in PVS-V, under a $-0.012~\rm{mmHg/mm}$ pressure gradient, the velocity is $4~\rm{\mu m/s}$, flowing from the intraorbital region toward the intracanalicular space. These results agree well with previous experimental and computational studies \cite{2022Arterial, boster2023artificial}.
	
	These findings emphasize the critical role of perivascular flow, glial-mediated transport, and osmotic gradients in regulating fluid homeostasis and metabolic waste clearance, reinforcing the glymphatic system’s function in the optic nerve. We emphasize that the interplay of biological structure, biophysical transporters and channels, and inescapable physical forces, is fundamentally the same as occurs in many cells, tissues, and organs in animals \cite{2020ATridomain, 2023Structural}.
	\subsection{Stimulus-Induced Fluid Circulation}
	
	To understand fluid dynamics during and after neural activation, we investigate the glymphatic system’s role, including glial cells and perivascular spaces, in generating stimulus-driven flow patterns. Neuronal activity disturbs the equilibrium state, inducing ionic redistribution and osmotic pressure gradients, which in turn drive fluid movement across compartments. Neuronal activity varies during the sleep cycle and that surely induces glymphatic flow. The variation in the frequency and location of action potentials during the sleep cycle is certain to produce ionic redistribution and osmotic pressure gradients, which in turn drive fluid movement across compartments \cite{2024Neuronal}. It is worthwhile to speculate that action potentials might produce other effects, e.g., dreams, beyond fluid flow.

	Figure~\ref{fig:Op_structure_model} left delineates stimulated and non-stimulated regions within the optic nerve ($\Omega_{OP}$). The stimulus, applied at $z=z_0$, mimics physiological neural activation. The applied current has a frequency of $50\ \rm{Hz}$ ($T=0.02\ \rm{s}$) with a duration of $0.2\ \rm{s}$ and a strength of $I_{sti}=3\times10^{-3}\ \rm{A/m^{2}}$ for $3\ \rm{ms}$.
	
	\begin{figure}[htb]
		\centering
		\includegraphics[width=0.8\linewidth]{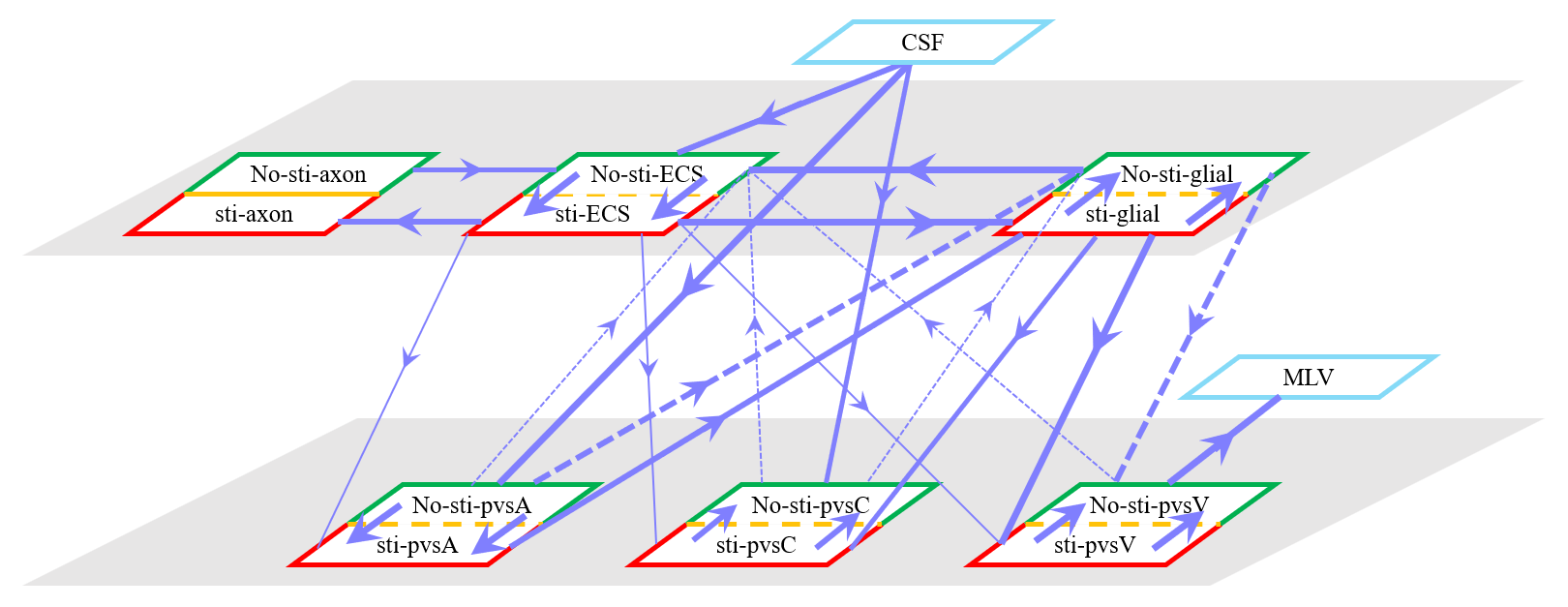}
		\caption{\label{fig:TransFluidFlux} Schematic of fluid flux between stimulated (red) and non-stimulated (green) regions, including trans-domain fluid exchange. Line thickness represents flux magnitude.}
	\end{figure}
	
	Figure~\ref{fig:TransFluidFlux} summarizes fluid movement across compartments, with two primary driving forces:
	\begin{itemize}
		\item Hydrostatic Pressure Gradients 
		\begin{itemize}
			\item     Higher intracranial pressure drives CSF from the SAS into the ECS, and PVS-A.
			\item In PVS-A, fluid flows from the intracranial region to the intraorbital region, while in PVS-V, it flows in the opposite direction. 
			\item A portion of CSF and interstitial fluid is cleared through the meningeal lymphatic vessels, which serve as a drainage route for metabolic waste and immune molecules. Fluid exits from PVS-V and SAS via meningeal lymphatics, ultimately draining into the deep cervical lymph nodes \cite{2017Evidence, 2020Glymphatic}.  
			\item The volume fractions of PVS-A and PVS-V decrease due to the drainage (Figure~\ref{fig:VolumFraction}d\&f).  
		\end{itemize}
		\item    Osmotic Pressure Gradients 
		\begin{itemize}
			\item Ionic redistribution during stimulation alters osmotic pressure, affecting trans-domain flux. 
			\item Decreased osmotic pressure in the ECS causes fluid leakage into glial compartments and perivascular spaces, reducing ECS volume (Figure~\ref{fig:VolumFraction}a) and increading Glial volume (Figure~\ref{fig:VolumFraction}b).  
			\item Fluid then moves through glial connexins toward the non-stimulated region, re-entering the ECS through AQP4 channels, establishing recirculation.
		\end{itemize}
	\end{itemize} 
	
	\begin{figure}
		\centering
		\includegraphics[width=\linewidth]{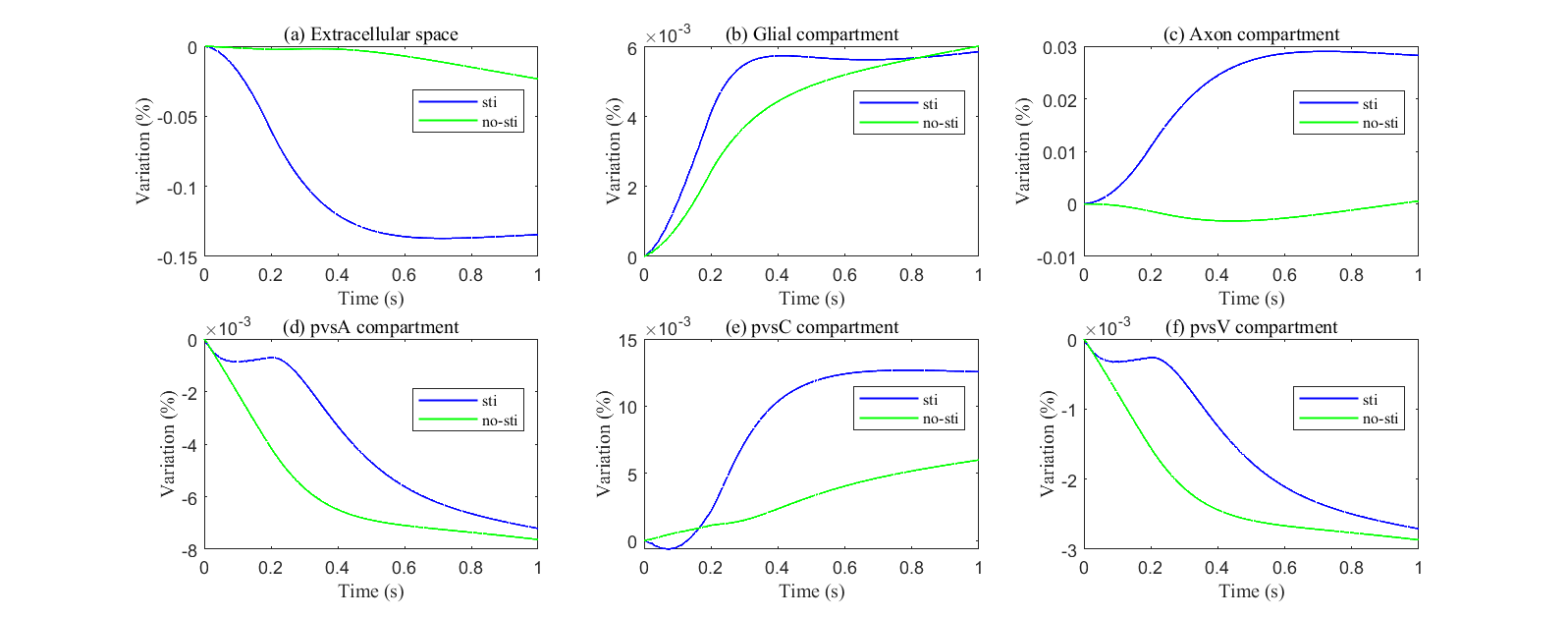}
		\caption{Variations of different compartments volume fractions during and after stimuli. }
		\label{fig:VolumFraction}
	\end{figure}

	Fluid circulation in the optic nerve is dynamically regulated by hydrostatic and osmotic pressure gradients, responding to neuronal activity through ion-driven osmotic effects. The glymphatic system facilitates waste clearance and volume regulation, ensuring homeostasis between stimulated and non-stimulated regions. These findings highlight the interconnected nature of CSF, glial, and perivascular flow, offering insights into the optic nerve's role in CNS fluid dynamics: interactions dominate as in most electro-osmotic systems. The structures of the nervous system link these flows; the physics of flow in the CSF, glia, and perivascular spaces depend on the interacting concentrations of the same salts. Everything interacts with everything else in such systems. Simplified models may suffice in special cases, but it is nearly impossible to define those special cases without a study of the full system, starting with conservation laws  and biological structure \cite{2023Structural}.
	
	\section{Parametric Study under Pathophysiological Conditions}
	Aquaporin (AQP) water channels, particularly Aquaporin-4 (AQP4), are highly expressed on astrocytes and play a crucial role in maintaining homeostasis in the central nervous system (CNS). 
	AQP4 expression is often dysregulated, leading to impaired glymphatic function and the accumulation of toxic metabolites \cite{rasmussen2018glymphatic}. In addition, changes in perivascular space permeability, which can result from metabolic waste blockage or structural alterations, further disrupt fluid dynamics and waste clearance \cite{2022Perivascular}. These pathological changes highlight the importance of understanding how variations in AQP4 expression and perivascular space permeability impact glymphatic function.
	
	In this section, we conduct a parametric study to investigate the effects of varying the trans-domain hydrostatic permeability of the glial membrane (reflecting AQP4 expression levels) and the intracompartment permeability of the perivascular space on fluid dynamics in the optic nerve.  By systematically varying these parameters, we aim to quantify their effects on fluid circulation, perivascular clearance, and pressure gradients, providing insights into the pathophysiological mechanisms underlying impaired glymphatic function.
	
	\subsection{Expression of Aquaporin (AQP) Water Channels}
	
	In our simulations, the hydrostatic permeability of the glial membrane, $L_{gl,l}$, for $l=pa,pc,pv,ex$, was varied by factors of $10^{-1}$, $10^{-2}$, and $10^{-4}$ to assess the impact of impaired AQP4 water channels on the glial membrane.
	
	As hydrostatic permeability decreases, the fluid trans-domain velocity across the glial membrane in the stimulated region is reduced, which leads to a decrease in total flux (see Fig. \ref{fig:TranRWater41}f and Fig. \ref{fig:TranWater41}a-c). Consequently, the radial fluid velocity within both the glial compartment and the ECS declines (see Fig. \ref{fig:TranRWater41}a-b). This results in a buildup of pressure in the stimulated region in both compartments (see Fig. \ref{fig:P_Phi_ex} and Fig.\ref{fig:P_O_gl}), which increases longitudinal fluid velocity (see Fig. \ref{fig:WatVecZdir}a \& c). The restriction in radial flux also reduces convection within the glial compartment and ECS.
	
	Under baseline conditions, fluid recirculates from the non-stimulated region back into the stimulated region (see Fig. \ref{fig:TransFluidFlux}), carrying sodium ($\rm{Na}^+$) ions along with it (see Fig. \ref{fig:NaDiffusion} in the Appendix). However, as hydrostatic permeability decreases, the rising pressure in the stimulated region further restricts this fluid return, leading to a reduction in radial fluid velocity. Consequently, the volume of the stimulated region decreases (see Fig. \ref{fig:ConVolu41}e), accompanied by a decline in sodium ($\rm{Na}^+$) and chloride ($\rm{Cl}^{-}$) concentrations (see Fig. \ref{fig:ConVolu41}b-c). This reduction in ionic concentrations lowers the osmotic pressure in the ECS following the stimulus. Meanwhile, the buildup of hydrostatic pressure increases trans-domain fluid velocities across perivascular space membranes (see Fig. \ref{fig:TranWater41}d-f), as fluid is primarily transported through the gaps between astrocytic endfeet rather than via AQP4 water channels. 
	
	
	\begin{figure}
		\centering
		\includegraphics[width=1\linewidth]{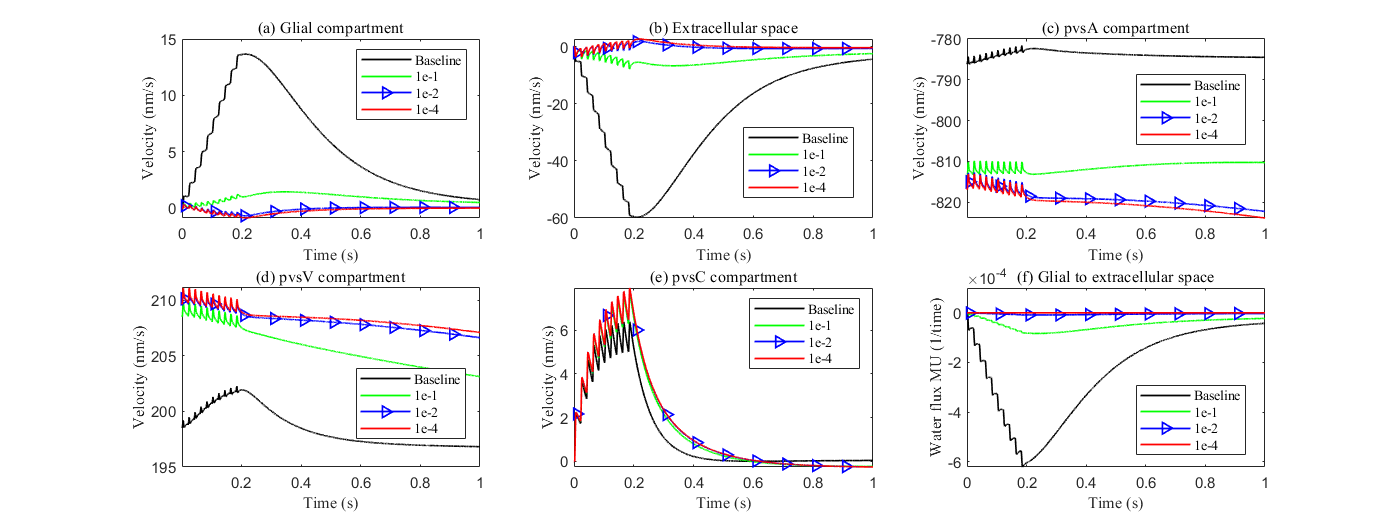}
		\caption{\label{fig:TranRWater41}   Average radial fluid velocity of  
			(a) Glial compartment; (b) ECS; (C) PVS-A; (d) PVS-V; (e) PVS-C; in the intradomain with varying levels of the hydrostatic permeability on the glial membrane; (f)  Average trans-domain fluid flux in the stimulated region.}
	\end{figure}
	\begin{figure}
		\centering
		\includegraphics[width=1\linewidth]{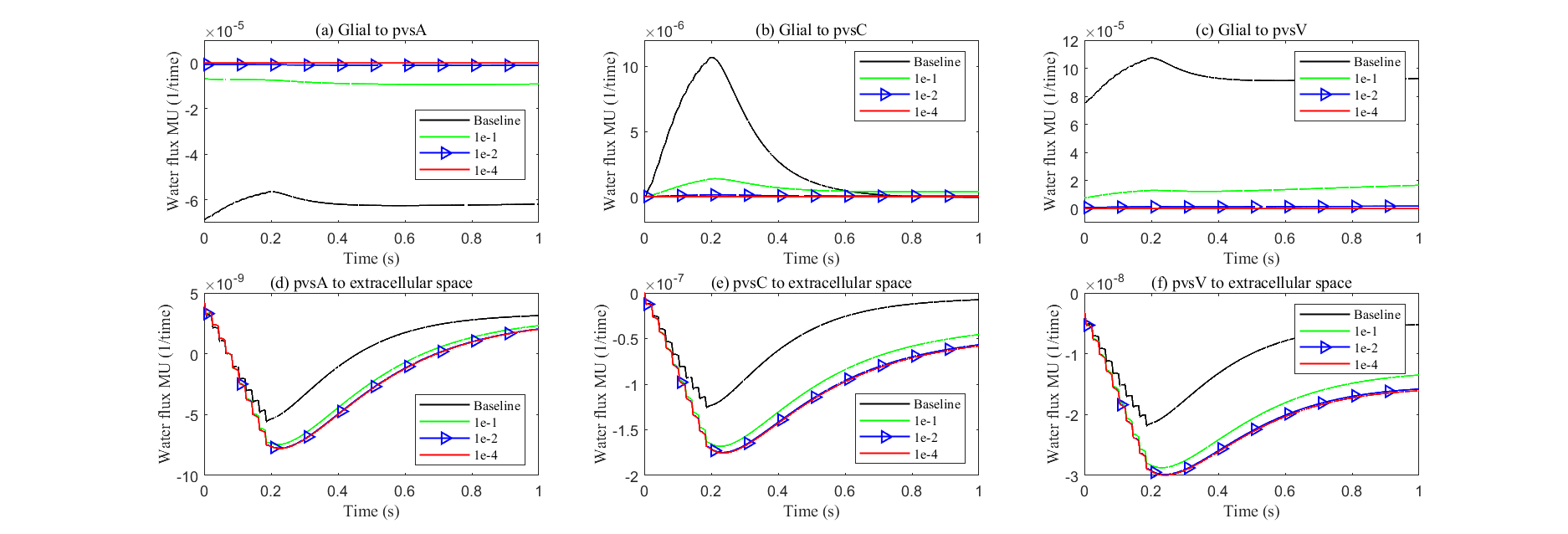}
		\caption{\label{fig:TranWater41}  Average trans-domain fluid flux  on perivascular spaces with varying levels of the hydrostatic permeability on the glial membrane in the stimulated region.}
	\end{figure}

	\begin{figure}
		\centering
		\includegraphics[width=1\linewidth]{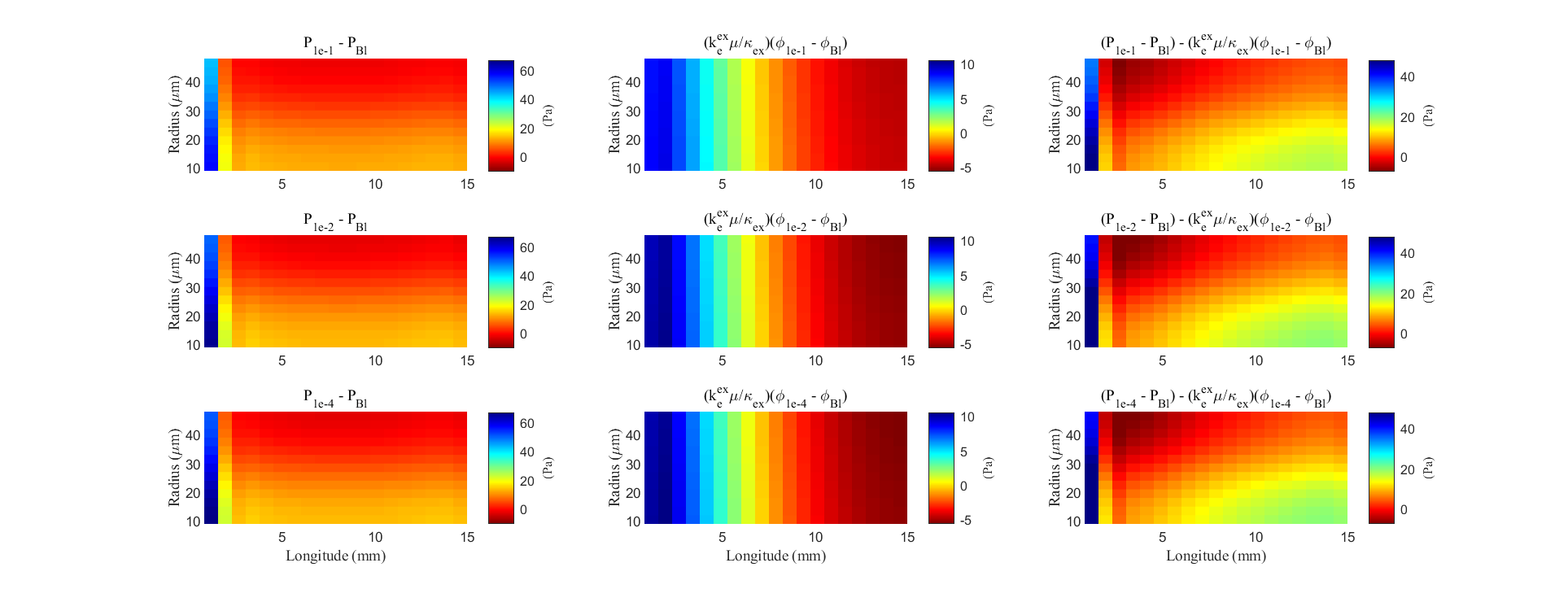}
		\caption{\label{fig:P_Phi_ex} The Spatial distribution of the difference of the hydrostatic pressure and the field potential within the ECS  with varying levels of the hydrostatic permeability on the glial membrane. "Bl" means the Baseline case.}
	\end{figure}
	
	\begin{figure}
		\centering
		\includegraphics[width=1\linewidth]{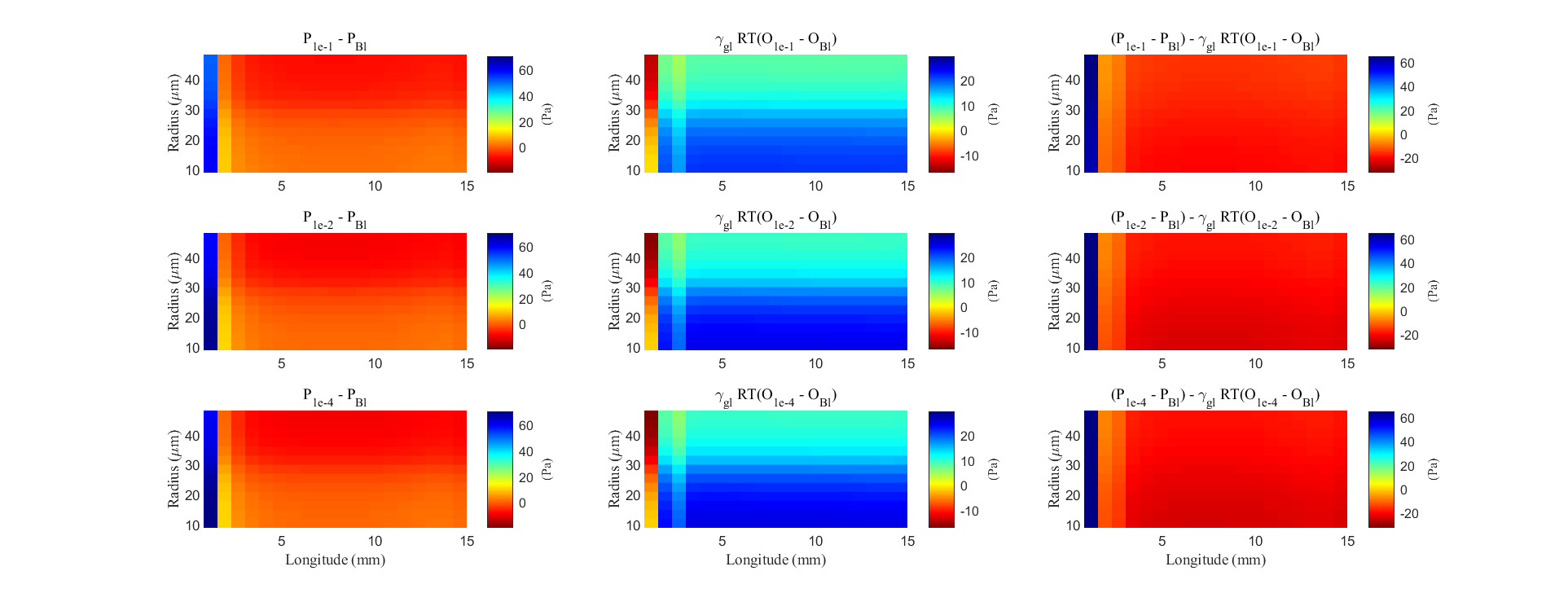}
		\caption{\label{fig:P_O_gl} The Spatial distribution of the difference of the hydrostatic pressure and the osmotic pressure within the glial compartment with varying levels of the hydrostatic permeability on the glial membrane. "Bl" means the Baseline case.}
	\end{figure}

	\begin{figure}
		\centering
		\includegraphics[width=1\linewidth]{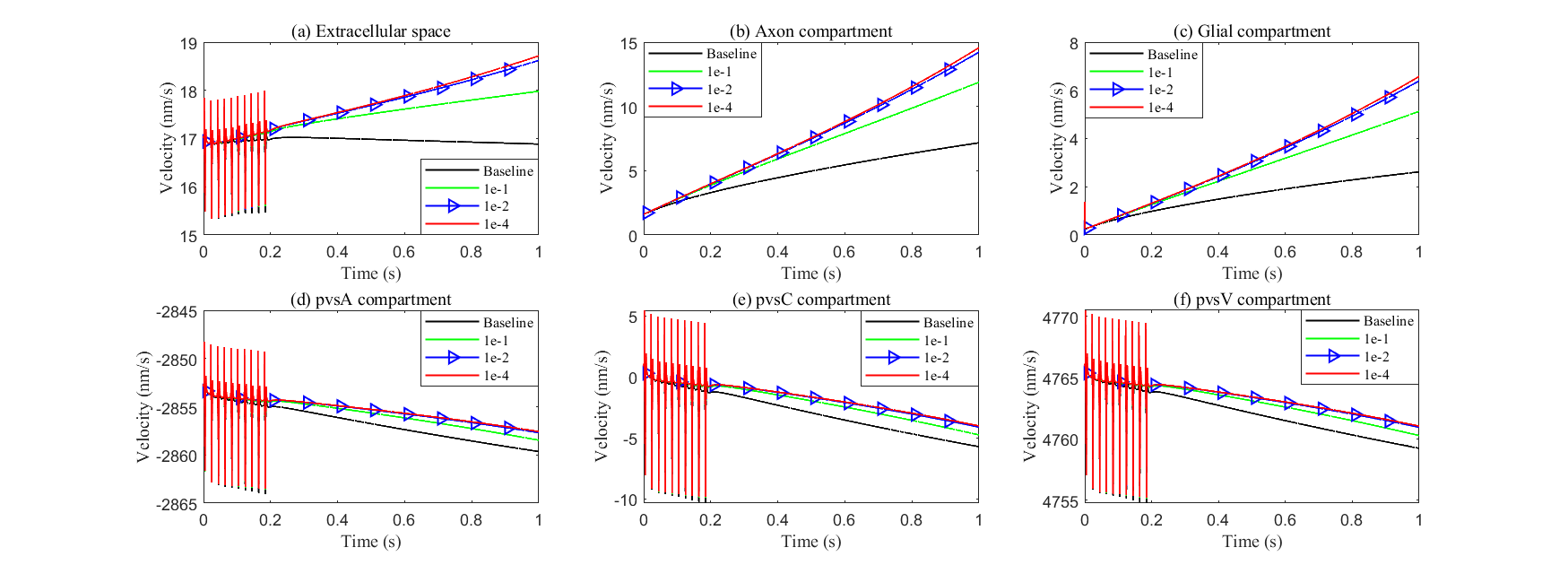}
		\caption{\label{fig:WatVecZdir}   Average longitude fluid velocity in the intradomain with varying levels of the hydrostatic permeability on the glial membrane.  (a) ECS; (b) Axon; (C) Glial; (d)-(f) PVS- A/C/V. }
	\end{figure}


	\begin{figure}
		\centering
		\includegraphics[width=1\linewidth]{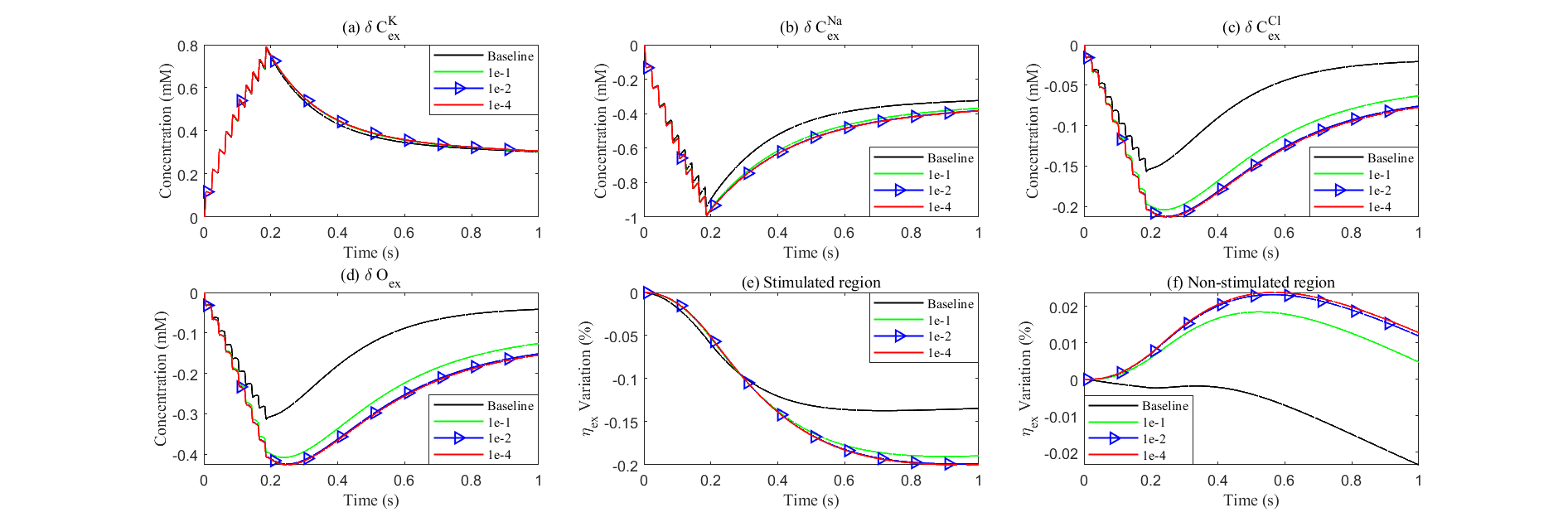}
		\caption{\label{fig:ConVolu41}  Variation of concentration and volume with varying levels of the hydrostatic permeability on the glial membrane. a-d: Average concentration variations inside the ECS  in the stimulated region. e-f: Average volume fraction variations in the ECS. Different lines denote results with varying hydrostatic permeability levels on the glial membrane.}
	\end{figure}

	\subsection{Functions of the Perivascular Space}
	In our previous studies \cite{2020ATridomain, 2019ABidomain}, the extracellular space and the perivascular spaces were treated as equivalent, without distinguishing between them.
	In this subsection, we examine  the effect of perivascular space on the flow by varying the transmebrane and intracompartment permeability.  
	
	\begin{itemize}
		\item \textbf{Baseline model:} Uses the parameters from Sections 3.
		
		\item \textbf{Case 1:} Increases the trans-domain exchange coefficient between the extracellular space and the perivascular spaces by a factor of $10^4$, based on the baseline parameters.
		
		\item \textbf{Case 2:} Adjusts the permeability within the perivascular spaces to match that of the extracellular space, while keeping the other parameters consistent with the baseline model.
		
		\item \textbf{Case 3:} Based on Case 2, sets the communication parameters between the cerebrospinal fluid and the perivascular spaces to match those of the extracellular space in the pia mater, with $\Gamma_{7}$ and $L_{pia,pa/pv} = L_{pia,ex}$.
	\end{itemize}

    Figure~\ref{fig:TransVelocitySti} summarizes the trans-domain fluid fluxes in the stimulated region for the four parameter cases. The directions of (i) ECS$\rightarrow$PVS exchange, (ii) glia$\rightarrow$PVS-V, and (iii) PVS-A$\rightarrow$glia are preserved across all cases; the only qualitative change occurs at the glia–PVS-C interface. Because the left, right, and bottom boundaries of PVS-C are no-flux, trans-domain exchange is the dominant pathway for volume adjustment within this compartment. In \textbf{Case~1} (enhanced ECS$\leftrightarrow$PVS exchange with relatively low intracompartmental permeability of PVS-C), flux proceeds primarily from the ECS into PVS-C, which then drives net transfer from PVS-C into the glial compartment. By contrast, in \textbf{Cases~2} and \textbf{3} the increased intracompartmental connectivity of PVS-C accelerates radial transport (Fig.~\ref{fig:SeveralVelocityR}e) and reverses the effective trans-domain direction: fluid exits the glial compartment and is conveyed toward the venous perivascular network.

Figure~\ref{fig:SeveralVelocityR} reports the space–averaged radial velocities in each compartment for the four parameter cases. 
\textbf{Extracellular space (ECS).} Relative to the baseline, increasing the ECS$\leftrightarrow$PVS trans-domain exchange coefficient (Case~1) drives more fluid from the ECS into PVS~A/C/V within the stimulated region (Fig.~\ref{fig:TransVelocitySti}d–f), thereby elevating the ECS radial velocity. In Cases~2 and~3, the time courses of the ECS$\leftrightarrow$PVS~A/V trans-domain fluxes remain similar to Case~1; however, the higher \emph{intracompartment} permeability of PVS~C facilitates lateral transport within PVS~C, increasing the net ECS$\to$PVS~C transfer (Fig.~\ref{fig:TransVelocitySti}e) and further augmenting the ECS radial velocity.

\textbf{Perivascular space A (PVS-A).} Enhancing the trans-domain exchange coefficient increases inflow into PVS-A but diverts volume through the membrane rather than within-plane, which reduces the radial velocity in Case~1. In Case~2, a reduction in the inner permeability of PVS-A further suppresses radial velocity relative to Case~1. In Case~3, weakened coupling between CSF and PVS-A yields very low radial velocity and promotes longitudinal redirection of fluid from the stimulated toward the non-stimulated region during stimulation.

\textbf{Perivascular spaces C and V (PVS-C/V).} In Case~3, a similar pattern emerges: diminished intracompartment connectivity and boundary coupling lower radial velocities, with flow preferentially redirected along the longitudinal axis rather than radially within the compartments.

	\begin{figure}
		\centering
		\includegraphics[width=1\linewidth]{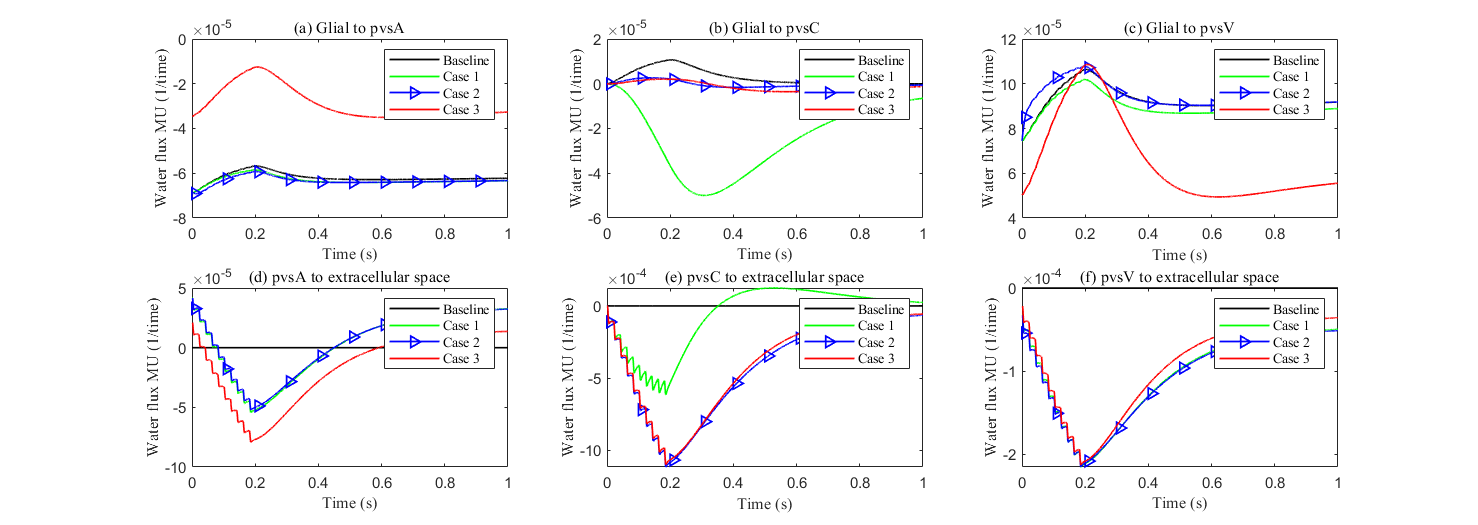}
		\caption{\label{fig:TransVelocitySti} The trans-domain fluid flux in the stimulated region with different PVS setup. (a-c) Glial to PVS- A/C/V; (c-f) PVS- A/C/V to ECS.  }
	\end{figure}
	
	\begin{figure}
		\centering
		\includegraphics[width=1\linewidth]{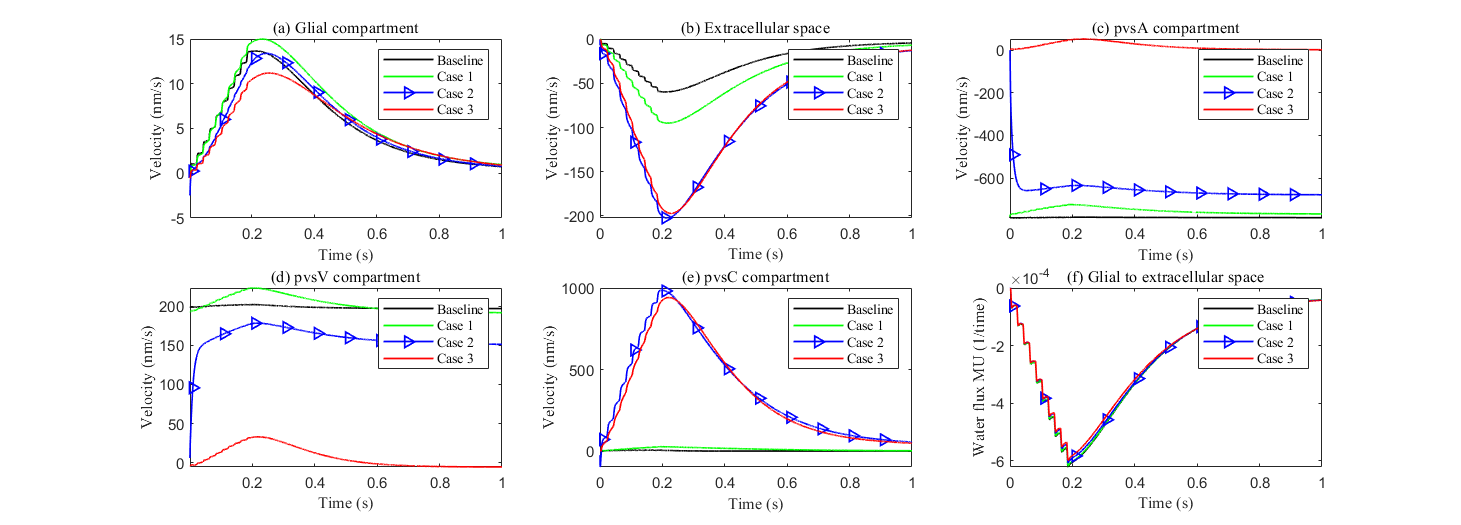}
		\caption{\label{fig:SeveralVelocityR}  Compartment radial velocity with different  PVS setup. (a) Glial; (b) ECS; (c-e) PVS- A/V/C. (f): The trans-domain fluid flux in the stimulated region.}
	\end{figure}

	\section{Metabolic Waste Clearance Mechanisms}
	
	With the deepening understanding of neurodegenerative diseases, studies have revealed that the accumulation of metabolic waste in the visual system is involved in neuronal degenerative changes, such as Alzheimer's disease \cite{1995Generation,1993Amyloid}, glaucoma, and optic atrophy. The progression of these diseases is often accompanied by the failure of metabolic clearance mechanisms. The perivascular spaces play a crucial role not only in maintaining water and ion balance in the visual system but also in the clearance of metabolic waste \cite{2021Perivascular,2022Electrostatic}. Dysfunction of these spaces may lead to the accumulation of metabolic waste and toxic substances in the extracellular space, further impairing neuronal health and exacerbating the pathological process. Study of the perivascular space seems an essential component of studying the brain blood barrier, and our study brings mathematical and physical specifity to that component of the overall biology and pathophysiology. Understanding the production and clearance mechanisms of metabolic waste in the visual system, as well as their relationship with diseases, will not only help uncover the pathological mechanisms of neurodegenerative diseases but also provide important clues for the development of targeted therapeutic strategies.
	
	\subsection{Model for Neutral Species}
	
	Compared to traditional ion homeostasis, the clearance of metabolic waste in the optic system involves water flow, as well as ion migration and diffusion. All occur in a complex structure with spatially nonuniform distribution of pumps and channels. The clearance of metabolic waste from the blood stream is done by similar mechanisms in the kidney. It is natural to imagine that the same physical mechanisms are used in the clearance of waste in the optic nerve, and indeed the membrane. Indeed, these mechanisms are fundamental properties of ionic solutions in all situations and so must exist in the brain and optic nerve. The question is not whether they exist but rather how does evolution exploit them. 

	In this study, we consider a simplified model  \cite{2019ABidomain,2021Optic,2020ATridomain} in which metabolic waste primarily enters the optic system. It lasts for 1 second from the left end, the ocular lens, and is cleared through two major extracellular pathways: (1) diffusion and convective transport within the extracellular space (ECS), and (2) drainage through gaps between glial endfeet into the PVS \cite{2016Glymphatic,2017Evidence}.
	
	Given the structural and functional constraints of the optic nerve microenvironment, our model explicitly accounts for the accumulation of metabolic waste in the extracellular space, where it undergoes diffusion, convection driven by bulk fluid flow, and clearance through the perivascular space. Solutes cannot pass through AQP4 channels and are therefore transported via advection along with solvent movement. When specific transport mechanisms for certain solutes, particularly critical metabolic waste, are discovered in the future, they can be readily incorporated into an extended version of our model once sufficiently characterized by biophysical measurements, including on isolated systems and preparations. 
    
    By integrating the known mechanisms, we simulate the microcirculation of metabolic waste as follows.
	
	\begin{subequations}\label{eq:Iontransport_r}
		\begin{align}
			& \dfrac{\partial }{\partial t}(\eta_{pa}C_{pa}^{Me})+\mathcal{M}_{pa,ex}J_{pa,ex}^{Me}+\nabla\cdot(\eta_{pa}\mathbf{j}_{pa}^{Me})=0,\\
			& \dfrac{\partial }{\partial t}(\eta_{pv}C_{pv}^{Me})+\mathcal{M}_{pv,ex}J_{pv,ex}^{Me}+\nabla\cdot(\eta_{pv}\mathbf{j}_{pv}^{Me})=0,\\
			& \dfrac{\partial }{\partial t}(\eta_{pc}C_{pc}^{Me})+\mathcal{M}_{pc,ex}J_{pc,ex}^{Me}+\nabla\cdot(\eta_{pc}\mathbf{j}_{pc}^{Me})=0,\\
			& \dfrac{\partial }{\partial t}(\eta_{ex}C_{ex}^{Me})-\sum_{k=pa,pv,pc}\mathcal{M}_{k,ex}J_{k,ex}^{Me}+\nabla\cdot(\eta_{ex}\mathbf{j}_{ex}^{Me})=0,
		\end{align}
	\end{subequations}
	
	The flux of metabolic waste across the interface consists of two components ($J_{k,ex}^{Me}$): one is the diffusive flux across the interface $G_k^{Me}log\left(\dfrac{C_{ex}^{Me}}{C_{k}^{Me}}\right)$, determined by the concentration difference of metabolic waste on both sides of the interface between different components; the other is the solute flow accompanying water movement across the interface $C_{up,wind}^{Me}U_{k,ex}$. 
	$$J_{k,ex}^{Me}=C_{up,wind}^{Me}U_{k,ex}+G_k^{Me}log\left(\dfrac{C_{ex}^{Me}}{C_{k}^{Me}}\right).$$
	
	\noindent Where $G_k^{Me}$ represents the permeability of metabolic waste exchange across the interface.
	Metabolic waste enters through the left boundary of the extracellular space at a constant flux, and the boundary conditions for the entire system are as follows:
	
	\begin{equation}
		\begin{cases}
			\nabla C_{l}^{Me}\cdot\hat{\mathbf{n}}_{r}=0, l=ex,pc & \mbox{on} \ \Gamma_1,\\
			\nabla C_{l}^{Me}\cdot\hat{\mathbf{n}}_{r}=\lambda_{l,down}(C_{l}^{Me}-C_{l}^{Me,\infty}), l=pa,pv & \mbox{on} \ \Gamma_1,\\
			\nabla C_{l}^{Me}\cdot\hat{\mathbf{n}}_{z}=-\lambda_{l,right}(C_{l}^{Me}-C_{l}^{Me,\infty}), l=ex,pa,pv & \mbox{on} \ \Gamma_2,\\
			\nabla C_{pc}^{Me}\cdot\hat{\mathbf{n}}_{z}=0 & \mbox{on} \ \Gamma_2,\\
			j_{ex,z}^{Me}=j_{constant} & \mbox{on} \ \Gamma_6,\\
			\nabla C_{l}^{Me}\cdot\hat{\mathbf{n}}_{z}=\lambda_{l,right}(C_{l}^{Me}-C_{l}^{Me,\infty}), l=pa,pv & \mbox{on} \ \Gamma_6,\\
			\nabla C_{pc}^{Me}\cdot\hat{\mathbf{n}}_{z}=0 & \mbox{on} \ \Gamma_6,\\
			\nabla C_{pv}^{Me}\cdot\hat{\mathbf{n}}_{r}=-\lambda_{pv,upper}(C_{pv}^{Me}-C_{pv}^{Me,\infty}) & \mbox{on} \ \Gamma_7,\\
			\mathbf{j}_{pa}^{Me}\cdot\hat{\mathbf{n}}_{r}=-G_{pia}^{Me}log\left(\dfrac{C_{csf}^{Me}}{C_{pa}^{Me}}\right)+C_{up,wind}^{Me}u_{pa,csf}, l=pc,ex & \mbox{on} \ \Gamma_7,\\
			\mathbf{j}_{l}^{Me}\cdot\hat{\mathbf{n}}_{r}=-G_{pia}^{Me}log\left(\dfrac{C_{csf}^{Me}}{C_{l}^{Me}}\right), l=pc,ex & \mbox{on} \ \Gamma_7,\\
		\end{cases}
	\end{equation}
	where $C_{l}^{Me,\infty}$ denotes the concentration of metabolic waste outside the optic nerve region.
	
	\subsection{Size-Dependent Clearance Mechanisms}
	
	The predominant clearance pathway of a metabolite in neural tissue is dictated by the \emph{effective size} (and resulting transport coefficients) of the solutes. In this work, we consider two operational classes: (i) \emph{small solutes}, typified by amyloid-$\beta$ monomers (A$\beta_{40/42}$, $\sim$4~kDa), and (ii) \emph{moderate species}, including tau monomers ($\sim$37–46~kDa)). This stratification aligns transport physics with biological structure and anatomy: the narrow extracellular space (ECS) supports rapid diffusion of small solutes, whereas larger assemblies, whose diffusivity is orders of magnitude lower, rely on  PVS convection driven by glymphatic flow. The central nervous system requires all waste to be cleared if it is to function. It appears that evolution has discovered how to use different mechanisms expressed and realized in just one  fundamental tissue structure.
	
	For small solutes we assign a higher ECS diffusivity $D_{\mathrm{ECS}}$ and modest ECS-PVS exchange, capturing diffusion-dominated clearance. For moderate species, we assign a markedly reduced $D_{\mathrm{ECS}}$ and strengthen the ECS-PVS exchange terms that couple to PVS convection, reflecting their reliance on directed glymphatic flow. This parameterization preserves a common framework while allowing the dominant physics to shift with molecular size.

The simulated spatial distributions in Figs.~\ref{fig:Small_molecule_space} and \ref{fig:Large_molecule_space} highlight dependence of clearance on the size of solutes. For A$\beta$ (small class), ECS concentration gradients dissipate rapidly as soon as they enter the computational domain. The resulting broad spread within the ECS enlarges the effective contact area with the PVS, promoting exchange along much of the interface (Fig.~\ref{fig:Small_trans_space}a-f). By contrast, for tau monomers (moderate class), ECS diffusion alone is insufficient: concentration profiles decay slowly, and effective removal requires trans–endfoot transfer into the PVS followed by convective transport. Consequently, appreciable ECS--PVS trans-domain flux is confined to a narrow zone near the left entry (Fig.~\ref{fig:Small_trans_space}g-l). 

The space-averaged dynamics are summarized in Figs.~\ref{fig:Small_molecule_Trans} and \ref{fig:Large_molecule_Trans}, and the net efflux through the right ECS boundary is shown in Fig.~\ref{fig:fluxonrightecs}. In Figs.~\ref{fig:Percentage_Small}--\ref{fig:Percentage_Large}, we report the fraction of material cleared through the PVS and through the right ECS boundary, normalized by the total mass remaining in the domain. For small molecules, more than $95\%$ of the cumulative clearance occurs via the PVS within $4\,\mathrm{s}$, whereas for moderate molecules the same threshold requires more than $15\,\mathrm{s}$. These results establish that (i) moderate molecules clear more slowly than small molecules, and (ii) PVS-mediated export is the dominant route for both classes, thus showing the important role of PVS-mediated export in the blood brain barrier. Notably, for small molecules the rapid ECS diffusion enlarges the ECS--PVS contact region and thereby \emph{enhances} PVS clearance efficiency. Thus, even under identical boundary conditions, the dominant clearance pathway bifurcates with solute size, consistent with experimental observations \cite{hladky2022glymphatic}.

	\begin{figure}
		\centering
		\includegraphics[width=0.6\linewidth]{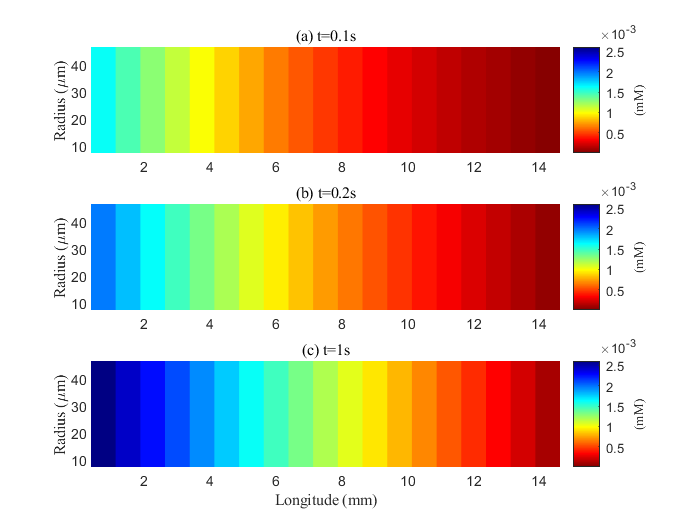}
		\caption{\label{fig:Small_molecule_space}  Spatial distribution of A$\beta$  in the ECS at different time points ($t=0.1s,0.2s,1s$). Effective clearance cannot be achieved within the extracellular space (ECS).}
	\end{figure}
	
	\begin{figure}
		\centering
		\includegraphics[width=0.6\linewidth]{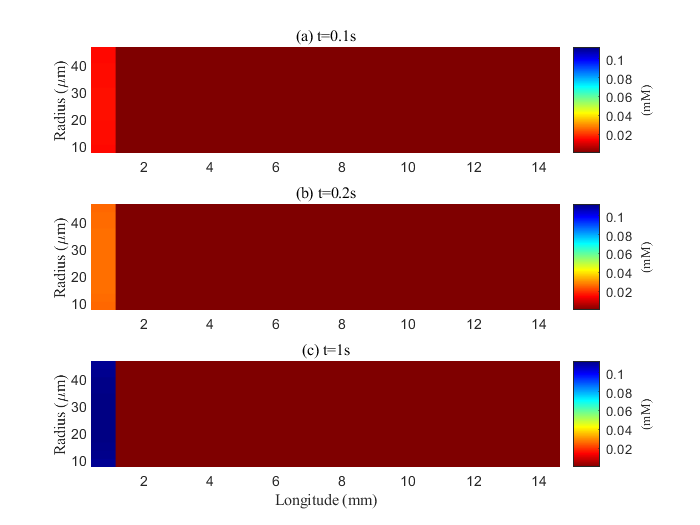}
		\caption{\label{fig:Large_molecule_space}  Spatial distribution of tau aggregates in the extracellular space at different time points ($t=0.1s,0.2s,1s$). Effective clearance via diffusion can occur within the narrow  extracellular space (ECS).}
	\end{figure}

	\subsection{Sleep and Wakefulness}

Sleep exerts a powerful influence on CNS metabolic homeostasis by enhancing the clearance of neurotoxic solutes \cite{2013sleep,2020Circadian}. A hallmark physiological change during sleep is the expansion of the extracellular space (ECS), which strengthens cerebrospinal fluid–interstitial fluid (CSF–ISF) exchange and increases the advective–diffusive transport capacity of the tissue. In parallel, glial (especially astrocytic) volume decreases, accommodating the enlarged ECS and improving intercompartment communication across astrocytic endfeet. Together, these shifts establish a fluid-dynamic milieu that favors solute removal.

To represent sleep in our model, we prescribe an increase in the ECS volume fraction \cite{2024Abrain}. Enlarging the ECS reduces geometric confinement, raises the effective mobility of interstitial fluid and small solutes, and widens the paracellular gaps between astrocytic endfeet, thereby increasing the effective contact area and exchange rate between ECS and PVS, particularly toward the PVS-V. Later versions of our model might allow changes in volume of compartments and thus seek the electro-osmotic causes for the volume changes, as well as compute them. 

The impact of sleep is strongly dependent on the size of solutes. For small solutes (e.g., A$\beta$), sleep-like ECS expansion markedly increases trans-domain flux across the ECS–PVS boundary and accelerates depletion from the ECS, as reflected in the spatial and temporal profiles in Figs.~\ref{fig:Small_molecule_Trans} and \ref{fig:Small_molecule_space}. In contrast, for moderate solutes (e.g., tau monomers), ECS diffusion remains slow and concentrations tend to accumulate near the left inflow region; although ECS enlargement modestly enhances exchange locally at the entry, the trans-domain flux is negligible across most of the domain (Fig.~\ref{fig:Small_trans_space}g-l), yielding only a tempered improvement in net clearance (Fig.~\ref{fig:Large_molecule_Trans}d).

Clearance fractions and timescales further underscore this selectivity dependent on solute size. The cumulative proportion of material exported via PVS versus the right ECS boundary (normalized by the in-domain mass) is shown in Figs.~\ref{fig:Percentage_Small}--\ref{fig:Percentage_Large}. For small solutes, more than $95\%$ of the clearance is accomplished through PVS within $\sim\!4\,\mathrm{s}$ under sleep-like conditions. For moderate solutes, the same PVS-dominated threshold is reached more slowly; increasing the ECS reduces the characteristic clearance time from approximately $20\,\mathrm{s}$ to $10\,\mathrm{s}$, indicating a meaningful benefit, although comparatively smaller. Overall, sleep shifts the transport landscape toward more efficient glymphatic export, with the greatest gains realized for rapidly diffusing solutes and a more modest, yet measurable, improvement for larger species that depend on PVS-guided convection following trans-endfoot transfer.

    \begin{figure}
		\centering
		\includegraphics[width=0.9\linewidth]{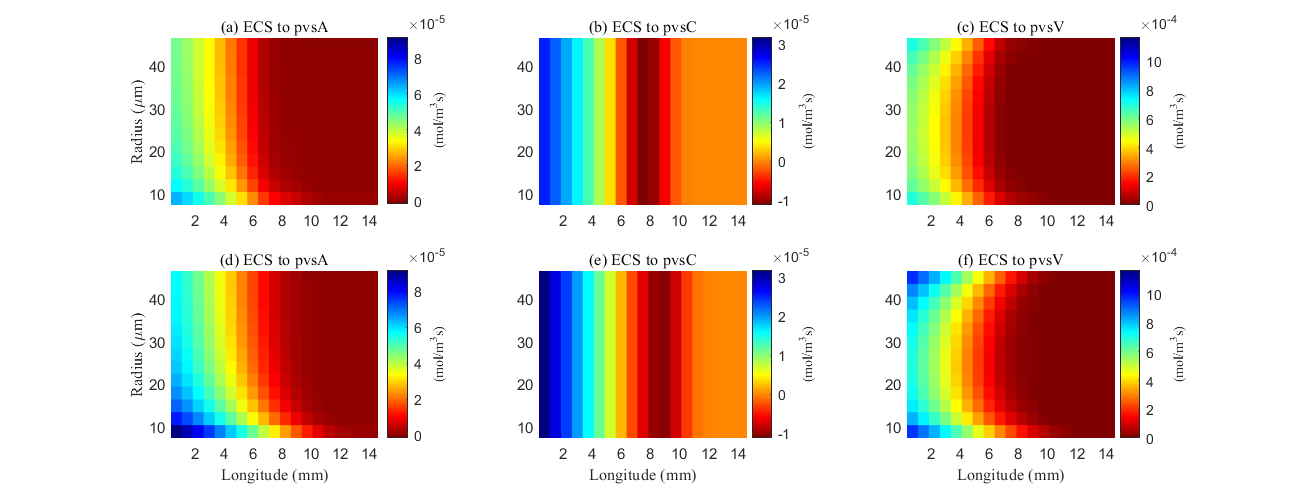}
		\includegraphics[width=0.9\linewidth]{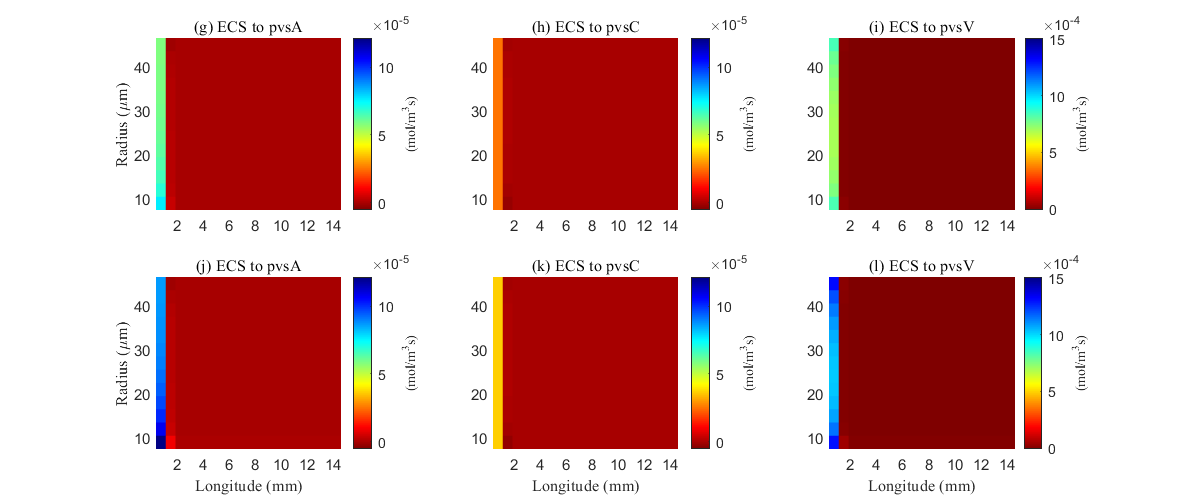}
		\caption{\label{fig:Small_trans_space} Space distribution of the trans-domain flux of A$\beta$ (a-f) and tau monomer (g-l) at time ($t=1s$). (a-c) ECS to PVS- A/C/V  during wakefulness states ($\eta$), and (d-f) ECS to PVS- A/C/V  during sleep period ($2\eta$). (g-i) ECS to PVS- A/C/V  during wakefulness states ($\eta$), and (j-l) ECS to PVS- A/C/V  during sleep period ($2\eta$).}
	\end{figure}

    
		\begin{figure}
		\centering
		\includegraphics[width=0.9\linewidth]{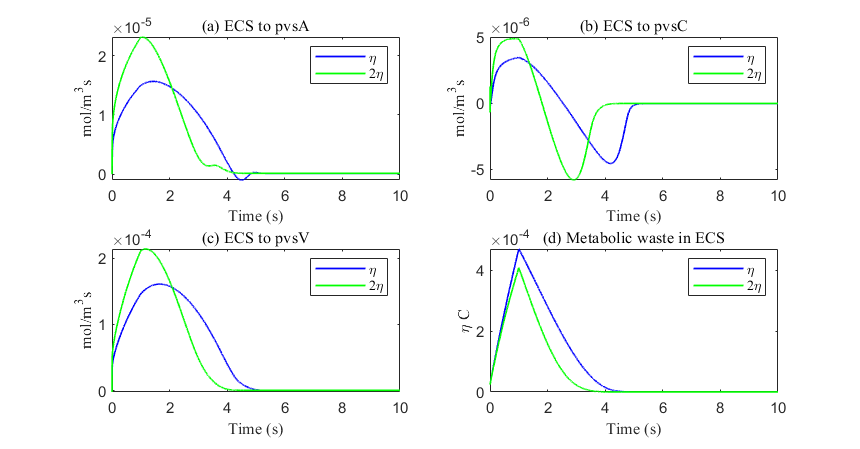}
		\caption{\label{fig:Small_molecule_Trans}   The trans-domain flux of A$\beta$  and the amount of A$\beta$ accumulated in ECS. (a) From ECS to PVS-A; (b) from ECS to PVS-C; (c) from ECS to PVS-V; (d) the amount of metabolic waste accumulated in ECS.}
	\end{figure}
    
	\begin{figure}
		\centering
		\includegraphics[width=0.9\linewidth]{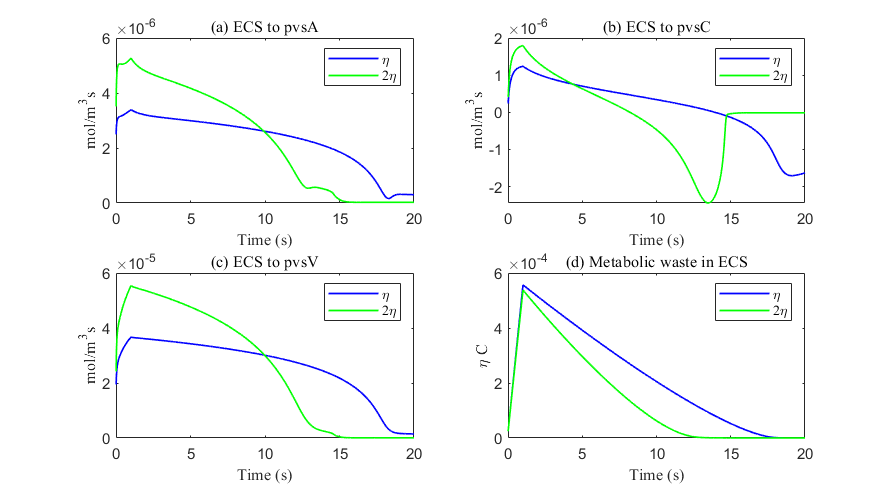}
		\caption{\label{fig:Large_molecule_Trans} The trans-domain flux of tau monomer  and the amount of tau monomer  accumulated in ECS. (a) From ECS to PVS-A; (b) from ECS to PVS-C; (c) from ECS to PVS-V; (d) the amount of metabolic waste accumulated in ECS.}
	\end{figure}

	\begin{figure} 
		\centering
		\includegraphics[width=0.4\linewidth]{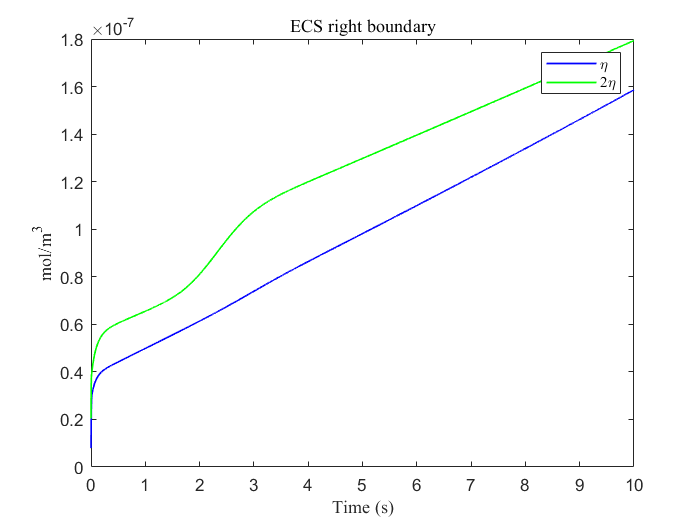}
        \includegraphics[width=0.4\linewidth]{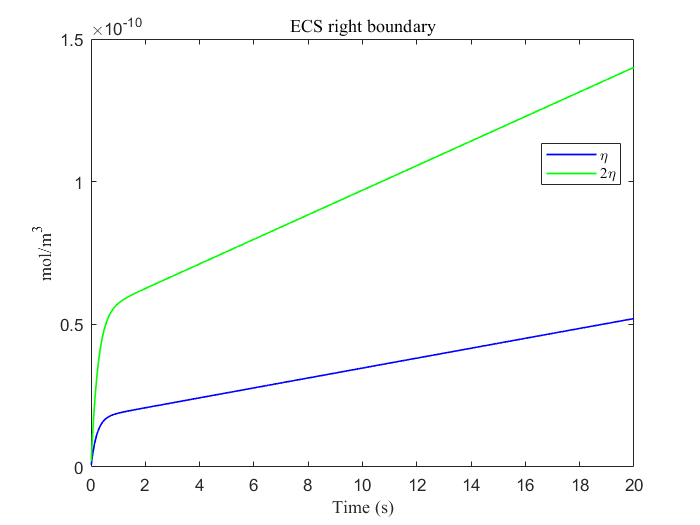}
		\caption{\label{fig:fluxonrightecs}  The amount of metabolic waste flux through the ECS right boundary accumulated over time. (Left)  A$\beta$; (Right) tau monomer.}
	\end{figure}
	
	\begin{figure}
		\centering
		\includegraphics[width=0.9\linewidth]{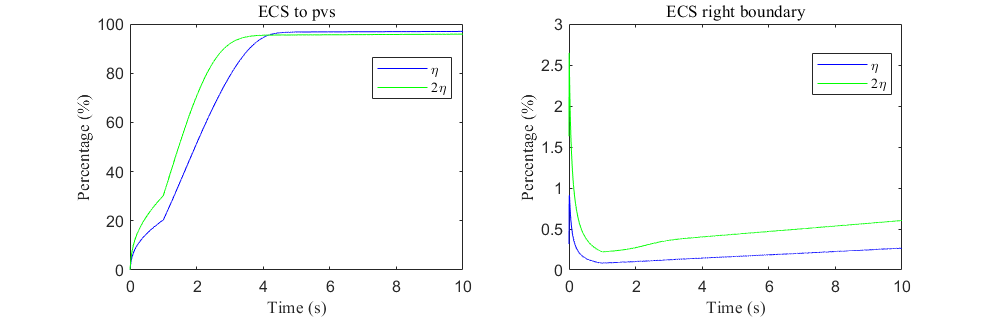}
		\caption{\label{fig:Percentage_Small} Clearance percentage of A$\beta$ through the perivascular spaces (Left) and the percentage cleared from within the ECS pathway via the right boundary (Right).}
	\end{figure}
	
	\begin{figure}
		\centering
		\includegraphics[width=0.9\linewidth]{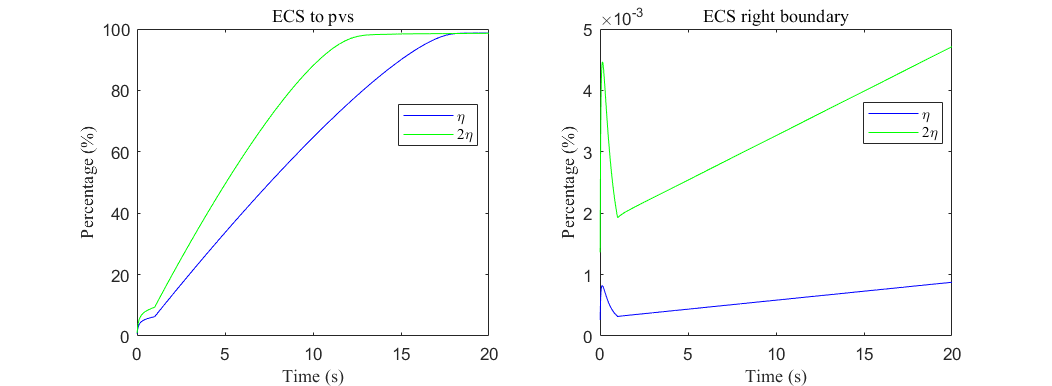}
		\caption{\label{fig:Percentage_Large} Clearance percentage of tau monomer through the perivascular spaces (Left) and the percentage cleared from within the ECS pathway via the right boundary (Right).}
	\end{figure}
	
        \subsection{Volume-Dependent Regulation of Trans-Domain Fluxes}
		
		In section 3.2, we provided a detailed examination of the specific processes of fluid circulation in the visual neural system during stimulation. As illustrated in Fig.~\ref{fig:TransFluidFlux}, fluid moves from the PVS-A into glial cells through the glial cell membrane, then from the glial cells into the PVS-V, and finally drains out of the brain via the MLV.
		
		In this system, driven by osmotic-hydrostatic pressure differences, fluid enters the glial cells through AQP4 channels located on the astrocytic endfeet, causing the endfeet to swell. This swelling compresses the gaps between adjacent endfeet, thereby reducing the permeability between the PVS-A and the ECS. Subsequently, fluid flows out from the intracellular space into the PVS-V, leading to the retraction of the endfeet. The expansion of the gaps between adjacent endfeet enhances the permeability between the ECS and the PVS-V. This dynamic process more closely approximates real biological conditions. To investigate the impact of this mechanism on the clearance efficiency of metabolic waste in the system, in this section, we study it by simply varying the permeability coefficients between the PVS-A, PVS-V, and the ECS. A more accurate mathematical model describing this dynamic change will be elaborated in our future research. In this section, we consider the clearance process of metabolic waste in the system during sleep in the previous section as the uniform case. For the non-uniform case, we simulated the expansion and contraction of endfoot gaps by modifying the permeability of PVS-A and ECS, and PVS-V and ECS, to represent glial cell swelling and contraction, respectively. For uniform case, $G_k^{2\eta,Me}=2G_k^{\eta,Me}$ and $L_{k,ex}^{2\eta}=2L_{k,ex}^{\eta}$, ($k=pa,pc,pv$). For non-uniform case, $G_{pa}^{2\eta,Me}=\frac{1}{4}G_{pa}^{\eta,Me}, G_{pv}^{2\eta,Me}=\frac{7}{4}G_{pv}^{\eta,Me}, G_{pc}^{2\eta,Me}=2G_{pc}^{\eta,Me}$, and $L_{pa,ex}^{2\eta}=\frac{1}{4}L_{pa,ex}^{\eta}, L_{pv,ex}^{2\eta}=\frac{7}{4}L_{pv,ex}^{\eta}, L_{pc,ex}^{2\eta}=2L_{pc,ex}^{\eta}$.
		
		As shown in the Fig. \ref{fig:pa_pv_7_1_percent_Small}, a comparison of metabolic waste clearance efficiency between the uniform and non-uniform cases reveals that more metabolic waste is cleared through the PVS-V pathway (Fig. \ref{fig:pa_pv_7_1_percent_Small}c), correspondingly reducing the amount cleared via the PVS-A pathway (Fig. \ref{fig:pa_pv_7_1_percent_Small}a). Overall, the non-uniform case demonstrates higher total clearance efficiency, as illustrated in Fig. \ref{fig:pa_pv_7_1_percent_Small}d. Fig. \ref{fig:pv_pa_7_1_Small} shows the clearance percentage of $A\beta$ through the perivascular sapces. The PVS-V represents an important pathway for metabolic waste clearance.
		
		\begin{figure}
			\centering
			\includegraphics[width=0.9\linewidth]{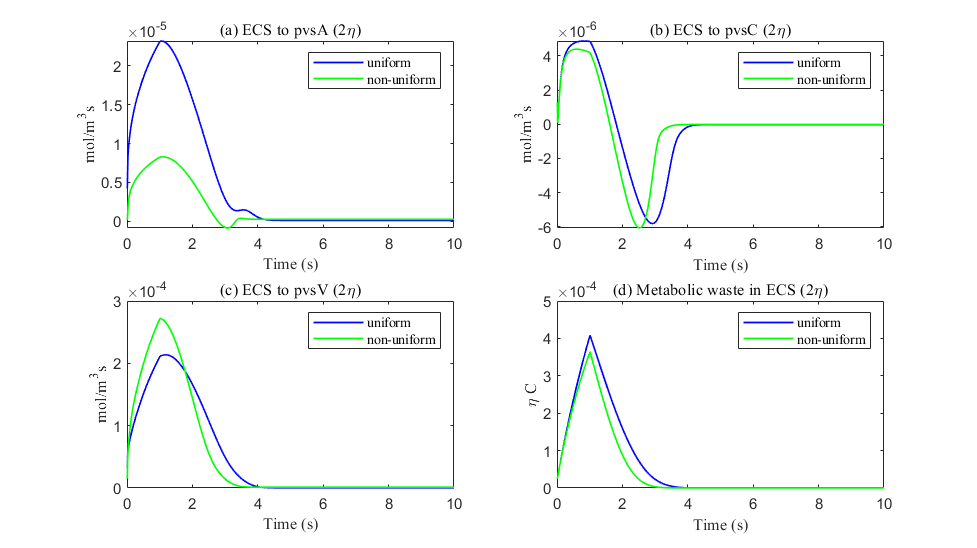}
			\caption{\label{fig:pa_pv_7_1_percent_Small} The trans-domain flux of tau monomer  and the amount of tau monomer  accumulated in ECS. (a) From ECS to PVS-A; (b) from ECS to PVS-C; (c) from ECS to PVS-V; (d) the amount of metabolic waste accumulated in ECS.}
		\end{figure}

		\begin{figure}
			\centering
			\includegraphics[width=0.9\linewidth]{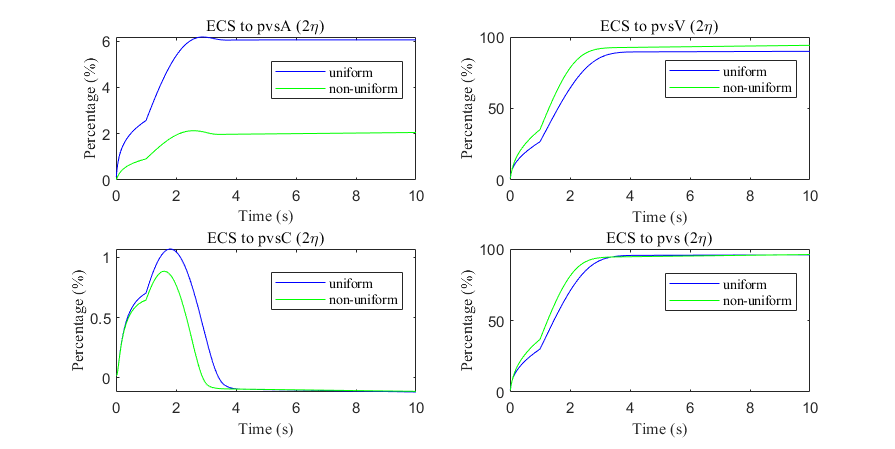}
			\caption{\label{fig:pv_pa_7_1_Small} The clearance percentage of $A\beta$ through the perivascular sapces.}
		\end{figure}
        
    \section{Conclusions}

We developed a multidomain, mechanistic model of optic–nerve microcirculation that couples hydrostatic and osmotic fluid transport with electro-diffusive solute movement across axons, glia, the extracellular space (ECS), and arterial/venous/capillary perivascular spaces (PVS-A/V/C). The model uses the well precedented electro-osmotic mechanisms found throughout animals and plants. The framework reproduces baseline glymphatic circulation, cerebrospinal fluid (CSF) entry from the subarachnoid space, exchange across astrocytic endfeet into ECS and glia, and reconvergence within PVS-V for drainage, and it captures stimulus-evoked responses in which ion-driven osmotic forces transiently reshape intercompartment flows. Parameter sweeps identify two decisive throughput controls: AQP4-mediated fluid permeability and PVS permeability. Reducing AQP4 suppresses radial exchange, elevates local pressure, and slows clearance, effects most pronounced for PVS-V–dependent species, whereas enhancing PVS connectivity restores directed convective removal.

Using this platform, we analyzed glymphatic clearance of metabolic wastes. A central finding is a size-dependent difference in kinetics and coupling, rather than a switch in the dominant pathway: across conditions, PVS-mediated export remains the primary route for both small and moderate solutes. Small molecules (e.g., A$\beta$ monomers) clear faster because rapid ECS diffusion spreads them broadly, enlarging the effective ECS–PVS contact region and thereby enhancing PVS exchange and downstream convection. In contrast, moderate/large species (e.g., tau monomers and small oligomers) have very low ECS diffusivity and depend more strictly on trans-endfoot transfer before being carried by PVS-V convection; their clearance is therefore slower and spatially localized near entry zones. This formulation reconciles the observed percentages, PVS-dominant export for both classes, with the size-dependent timescales and flux patterns.
The model also explains the sleep–wake effect mechanistically: enlarging ECS volume and permeability increases trans-domain flux and accelerates removal of small solutes, while yielding only modest gains for moderate-size species whose bulk trans-domain exchange remains limited.

There exist several limitations in this work.    The transport repertoire includes a restricted set of channels and transporters, and biochemical processing is simplified (e.g., receptor–mediated export and enzymatic degradation are not explicitly resolved). Inflammatory and oxidative pathophysiology, as well as explicit aggregation kinetics, are omitted, and the geometry/parameters are idealized for the optic nerve. Our representation of sleep is intentionally minimalist: the ECS volume–fraction increase is prescribed rather than emergent from coupled osmotic–mechanical dynamics. These choices improve tractability but limit quantitative generalization and mechanistic fidelity in some scenarios.

Future work will integrate aggregation kinetics (LLPS, primary/secondary nucleation, fragmentation) with transport to link residence time to misfolding risk; incorporate cellular uptake pathways and BBB export explicitly; and couple the optic–nerve microcirculation to meningeal lymphatic dynamics and venous pressure regulation. Moving toward subject–specific three–dimensional geometries with data–driven calibration and uncertainty quantification (e.g., machine–learning–assisted inference) should strengthen translational relevance to glaucoma, cerebral amyloid angiopathy, and Alzheimer’s disease, and guide interventions targeting AQP4 polarization, PVS patency, and sleep–dependent enhancement of clearance.

	\section*{Acknowledgments}
	This work was partially supported by the National Natural Science Foundation of China no. 12231004 (H. Huang) and 12071190 (S. Xu).

	\bibliographystyle{unsrt}
	\bibliography{sample}

\begin{thebibliography}{10}

\bibitem{2012AParavascular}
Jeffrey~J. Iliff, Minghuan Wang, Yonghong Liao, Benjamin~A. Plogg, and Weiguo
  Peng.
\newblock A paravascular pathway facilitates csf flow through the brain
  parenchyma and the clearance of interstitial solutes, including amyloid beta.
\newblock {\em Science translational medicine}, 4(147):147, 2012.

\bibitem{2015TheGlymphatic}
Nadia~Aalling Jessen, Anne Sofie~Finmann Munk, Iben Lundgaard, and Maiken
  Nedergaard.
\newblock The glymphatic system: A beginner's guide.
\newblock {\em Neurochemical Research}, 40(12):2583--2599, 2015.

\bibitem{2016Glymphatic}
Mahdi Asgari, Diane de~Zélicourt, and Vartan Kurtcuoglu.
\newblock Glymphatic solute transport does not require bulk flow.
\newblock {\em Scientific Reports}, 6(1):1--16, 2016.

\bibitem{2017Impairment}
Quan Jiang, Li~Zhang, Guangliang Ding, Esmaeil Davoodi-Bojd, Qingjiang Li, Lian
  Li, Neema~S Sadry, Michael Chopp, and Zhenggang Zhang.
\newblock Impairment of the glymphatic system after diabetes.
\newblock {\em SAGE Publications}, 37(4):1326--1337, 2017.

\bibitem{2020Glymphatic}
Maiken Nedergaard and Steven~A. Goldman.
\newblock Glymphatic failure as a final common pathway to dementia.
\newblock {\em Science}, 370(6512):50--56, 2020.

\bibitem{2024Modeling}
Marie E. Rognes et~al. Lars Willas~Dreyer, Anders~Eklund.
\newblock Modeling csf circulation and the glymphatic system during infusion
  using subject specific intracranial pressures and brain geometries.
\newblock {\em Fluids and Barriers of the CNS}, 1, 2024.

\bibitem{2018Flow}
Humberto Mestre, Jeffrey Tithof, Ting Du, Wei Song, and Douglas~H. Kelley.
\newblock Flow of cerebrospinal fluid is driven by arterial pulsations and is
  reduced in hypertension.
\newblock {\em Nature Communications}, 9(1):4878--4887, 2018.

\bibitem{2022Perivascular}
Benjamin~V. Ineichen, Serhat~V. Okar, Steven~T. Proulx, Britta Engelhardt, Hans
  Lassmann, and Daniel~S. Reich.
\newblock Perivascular spaces and their role in neuroinflammation.
\newblock {\em Neuron}, 110(21):3566--3581, 2022.

\bibitem{1992Alzheimer}
J.~A. Hardy and G.~A. Higgins.
\newblock Alzheimer's disease: the amyloid cascade hypothesis.
\newblock {\em Science}, 256(5054):184--185, 1992.

\bibitem{2016Theamyloid}
Dennis~J Selkoe and John Hardy.
\newblock The amyloid hypothesis of alzheimer's disease at 25years.
\newblock {\em Embo Molecular Medicine}, 8(6):595--608, 2016.

\bibitem{2013sleep}
Lulu Xie, Hongyi Kang, Qiwu Xu, Michael~J Chen, Yonghong Liao,
  Meenakshisundaram Thiyagarajan, John O’Donnell, Daniel~J Christensen,
  Charles Nicholson, Jeffrey~J Iliff, et~al.
\newblock Sleep drives metabolite clearance from the adult brain.
\newblock {\em science}, 342(6156):373--377, 2013.

\bibitem{fultz2019coupled}
Nina~E Fultz, Giorgio Bonmassar, Kawin Setsompop, Robert~A Stickgold, Bruce~R
  Rosen, Jonathan~R Polimeni, and Laura~D Lewis.
\newblock Coupled electrophysiological, hemodynamic, and cerebrospinal fluid
  oscillations in human sleep.
\newblock {\em Science}, 366(6465):628--631, 2019.

\bibitem{2024Neuronal}
Li~Feng Jiang-Xie, Antoine Drieu, Kesshni Bhasiin, Daniel Quintero, Igor
  Smirnov, and Jonathan Kipnis.
\newblock Neuronal dynamics direct cerebrospinal fluid perfusion and brain
  clearance.
\newblock {\em Nature}, (Mar.7 TN.8002):627, 2024.

\bibitem{2003Arterial}
Lynne~E. Bilston, David~F. Fletcher, Andrew~R. Brodbelt, and Marcus~A.
  Stoodley.
\newblock Arterial pulsation-driven cerebrospinal fluid flow in the
  perivascular space: A computational model.
\newblock {\em Computer Methods in Biomechanics $\&$ Biomedical Engineering},
  6(4):235--241, 2003.

\bibitem{boster2023artificial}
Kimberly~AS Boster, Shengze Cai, Antonio Ladr{\'o}n-de Guevara, Jiatong Sun,
  Xiaoning Zheng, Ting Du, John~H Thomas, Maiken Nedergaard, George~Em
  Karniadakis, and Douglas~H Kelley.
\newblock Artificial intelligence velocimetry reveals in vivo flow rates,
  pressure gradients, and shear stresses in murine perivascular flows.
\newblock {\em Proceedings of the National Academy of Sciences},
  120(14):e2217744120, 2023.

\bibitem{chou2024machine}
Dean Chou and Po-Yen Chen.
\newblock A machine learning method to explore the glymphatic system via
  poroelastodynamics.
\newblock {\em Chaos, Solitons \& Fractals}, 178:114334, 2024.

\bibitem{2022glymphatic}
Tomas Bohr, Poul~G Hjorth, Sebastian~C Holst, Sabina Hrab{\v{e}}tov{\'a}, Vesa
  Kiviniemi, Tuomas Lilius, Iben Lundgaard, Kent-Andre Mardal, Erik~A Martens,
  Yuki Mori, et~al.
\newblock The glymphatic system: Current understanding and modeling.
\newblock {\em IScience}, 25(9), 2022.

\bibitem{Feher2012Quantitative}
Feher and Joseph.
\newblock Quantitative human physiology.
\newblock {\em Elsevier/Academic Press}, 2012.

\bibitem{silverthorn2019human}
Dee~Unglaub Silverthorn, Bruce~R Johnson, William~C Ober, CE~Ober,
  A~Impagliazzo, and AC~Silverthorn.
\newblock {\em Human physiology: An integrated approach}.
\newblock Pearson San Francisco, CA, 2019.

\bibitem{2020ATridomain}
Yi~Zhu, Shixin Xu, Robert~S Eisenberg, and Huaxiong Huang.
\newblock A tridomain model for potassium clearance in optic nerve of necturus.
\newblock {\em Biophysical journal}, 120(15):3008--3027, 2021.

\bibitem{2023Structural}
R.~Eisenberg.
\newblock Structural analysis of fluid flow in complex biological systems.
\newblock {\em Modeling and Artificial Intelligence in Ophthalmology}, 2023.

\bibitem{1985Epithelial}
R~T Mathias.
\newblock Epithelial water transport in a balanced gradient system.
\newblock {\em Biophysical Journal}, 47(6):823--836, 1985.

\bibitem{1997Physiological}
R~T Mathias, J~L Rae, and G~J Baldo.
\newblock Physiological properties of the normal lens.
\newblock {\em Physiological Reviews}, 77(1):21--50, 1997.

\bibitem{2019ABidomain}
Yi~Zhu, Shixin Xu, Robert Eisenberg, and Huaxiong Huang.
\newblock A bidomain model for lens microcirculation.
\newblock {\em Biophysical journal}, 116(6):1171--1184, 2019.

\bibitem{2011Lens}
Junyuan Gao, Xiurong Sun, Leon~C Moore, Thomas~W White, and Richard~T Mathias.
\newblock Lens intracellular hydrostatic pressure is generated by the
  circulation of sodium and modulated by gap junction coupling.
\newblock {\em The Journal of General Physiology}, 137(6):507--520, 2011.

\bibitem{2020Quantitative}
Wenyu Deng, Crystal Liu, Carlos Parra, Jeffrey~R. Sims, and Kevin~C. Chan.
\newblock Quantitative imaging of the clearance systems in the eye and the
  brain.
\newblock {\em Quantitative Imaging in Medicine and Surgery}, 10(1):1--14,
  2020.

\bibitem{2009Ischemic}
Sohan~Singh Hayreh.
\newblock Ischemic optic neuropathy.
\newblock {\em Progress in retinal and eye research}, 28(1):34--62, 2009.

\bibitem{2017Evidence}
Mathieu Emily, Gupta Neeru, Ahari Amir, Zhou Xun, Hanna Joseph, and Yücel~Yeni
  H.
\newblock Evidence for cerebrospinal fluid entry into the optic nerve via a
  glymphatic pathway.
\newblock {\em Investigative Ophthalmology $\&$ Visual Science}, 58(11):4784,
  2017.

\bibitem{1966Physiological}
SW~Kuffler, JG~Nicholls, and RK~Orkand.
\newblock Physiological properties of glial cells in the central nervous system
  of amphibia.
\newblock {\em Journal of Neurophysiology}, 29(4):768--787, 1966.

\bibitem{1966Effect}
RK~Orkand, JG~Nicholls, and SW~Kuffler.
\newblock Effect of nerve impulses on the membrane potential of glial cells in
  the central nervous system of amphibia.
\newblock {\em Journal of neurophysiology}, 29(4):788--806, 1966.

\bibitem{2017Aquaporin}
Tsutomu Nakada, Ingrid Kwee, Hironaka L.Igarashi, and Yuji Suzuki.
\newblock Aquaporin-4 functionality and virchow-robin space water dynamics:
  Physiological model for neurovascular coupling and glymphatic flow.
\newblock {\em Journal of Turbulence}, 18(8), 2017.

\bibitem{2017Astrocytic}
Alba Bellot-Saez, Orsolya Kekesi, J.W. Morley, and Yossi Buskila.
\newblock Astrocytic modulation of neuronal excitability through k+ spatial
  buffering.
\newblock {\em Neuroscience and Biobehavioral Reviews}, 77:87--97, 2017.

\bibitem{2006Theimpact}
Anke, Wallraff, Rüdiger, Khling, Uwe, Heinemann, Martin, Theis, Klaus, and
  Willecke.
\newblock The impact of astrocytic gap junctional coupling on potassium
  buffering in the hippocampus.
\newblock {\em Journal of Neuroscience the Official Journal of the Society for
  Neuroscience}, 26(20):5438--5447, 2006.

\bibitem{2011Anaquaporin4}
Valentina Benfenati, Marco Caprini, Melania Dovizio, Maria~N Mylonakou, and
  Mahmood Amiry-Moghaddam.
\newblock An aquaporin-4/transient receptor potential vanilloid 4 (aqp4/trpv4)
  complex is essential for cell-volume control in astrocytes.
\newblock {\em Proceedings of the National Academy of Sciences},
  108(6):2563--2568, 2011.

\bibitem{2017Mechanisms}
Alejandra Daruich, Alexandre Matet, Alexandre Moulin, Laura Kowalczuk, Micha?L
  Nicolas, Alexandre Sellam, Pierre~Rapha?L Rothschild, Samy Omri, Emmanuelle
  Gélizé, and Laurent~and Jonet.
\newblock Mechanisms of macular edema: Beyond the surface.
\newblock {\em Progress in Retinal and Eye Research}, 63:20--68, 2017.

\bibitem{1984Thesheath}
Singh~Sohan Hayreh.
\newblock The sheath of the optic nerve.
\newblock {\em Ophthalmologica}, 189(1-2):54--63, 1984.

\bibitem{2003Anatomic}
J.~B. Jonas, E.~Berenshtein, and L.~Holbach.
\newblock Anatomic relationship between lamina cribrosa, intraocular space, and
  cerebrospinal fluid space.
\newblock {\em Investigative ophthalmology $\&$ visual science},
  44(12):5189--5195, 2003.

\bibitem{1999Transport}
W.~M. Lai and V.~C. Mow.
\newblock Transport of multi-electrolytes in charged hydrated biological soft
  tissues.
\newblock {\em Transport in Porous Media}, 34(1-3):143--157, 1999.

\bibitem{Sibille2015The}
Jérémie Sibille, Khanh Dao~Duc, David Holcman, Nathalie Rouach, and Renaud
  Jolivet.
\newblock The neuroglial potassium cycle during neurotransmission: Role of
  kir4.1 channels.
\newblock {\em Plos Computational Biology}, 11(3):e1004137, 2015.

\bibitem{2017Quantitative}
Joseph~J Feher.
\newblock {\em Quantitative Human Physiology: An Introduction}.
\newblock Academic press, 2017.

\bibitem{2018Osmosis}
Shixin Xu, Bob Eisenberg, Zilong Song, and Huaxiong Huang.
\newblock Osmosis through a semi-permeable membrane: a consistent approach to
  interactions.
\newblock {\em arXiv preprint arXiv}, 1806.00646, 2018.

\bibitem{Mori2015A}
Mori Yoichiro.
\newblock A multidomain model for ionic electrodiffusion and osmosis with an
  application to cortical spreading depression.
\newblock {\em Physica D Nonlinear Phenomena}, 308:94--108, 2015.

\bibitem{1985Steady}
R~T Mathias.
\newblock Steady-state voltages, ion fluxes, and volume regulation in syncytial
  tissues.
\newblock {\em Biophysical Journal}, 48(3):435--448, 1985.

\bibitem{1979Electrical}
R.T. Mathias, J.L. Rae, and R.S. Eisenberg.
\newblock Electrical properties of structural components of the crystalline
  lens.
\newblock {\em Biophysical Journal}, 25(1):181--201, 1979.

\bibitem{2014Self}
Li~Wan, Shixin Xu, Maijia Liao, Chun Liu, and Ping Sheng.
\newblock Self-consistent approach to global charge neutrality in
  electrokinetics: A surface potential trap model.
\newblock {\em Physical Review X}, 4(1):011042, 2014.

\bibitem{2012Development}
Ehsan Vaghefi, Duane T~K Malcolm, Marc~D Jacobs, and Paul~J Donaldson.
\newblock Development of a 3d finite element model of lens microcirculation.
\newblock {\em BioMedical Engineering OnLine,11,1(2012-09-19)}, 11(1):69, 2012.

\bibitem{1983Effect}
R~T Mathias.
\newblock Effect of tortuous extracellular pathways on resistance measurements.
\newblock {\em Biophysical Journal}, 42(1):55--59, 1983.

\bibitem{2001Diffusion}
Charles Nicholson.
\newblock Diffusion and related transport mechanisms in brain tissue.
\newblock {\em Reports on Progress in Physics}, 64(7):815, 2001.

\bibitem{1995Extracellular}
MA~Pérez-Pinzón, LIAN Tao, and CHARLES Nicholson.
\newblock Extracellular potassium, volume fraction, and tortuosity in rat
  hippocampal ca1, ca3, and cortical slices during ischemia.
\newblock {\em Journal of Neurophysiology}, 74(2):565--573, 1995.

\bibitem{1977Electrical}
R~T Mathias, R~S Eisenberg, and R.~Valdiosera.
\newblock Electrical properties of frog skeletal muscle fibers interpreted with
  a mesh model of the tubular system.
\newblock {\em Biophysical Journal}, 17(1):57--93, 1977.

\bibitem{1985Electro}
STUART McLAUGHLIN and RICHARD~T Mathias.
\newblock Electro-osmosis and the reabsorption of fluid in renal proximal
  tubules.
\newblock {\em The Journal of general physiology}, 85(5):699--728, 1985.

\bibitem{2009Intracellular}
Leah~R Band, Cameron~L Hall, Giles Richardson, and andAlexanderJEFoss
  Oliver~EJensen, JenniferHSiggers.
\newblock Intracellular flow in optic nerve axons: A mechanism for cell death
  in glaucoma.
\newblock {\em Investigative Opthalmology $\&$ Visual Science},
  50(8):3750--3758, 2009.

\bibitem{ray2021quantitative}
Lori~A Ray, Martin Pike, Matthew Simon, Jeffrey~J Iliff, and Jeffrey~J Heys.
\newblock Quantitative analysis of macroscopic solute transport in the murine
  brain.
\newblock {\em Fluids and Barriers of the CNS}, 18:1--19, 2021.

\bibitem{1960Thresholds}
Richard Fitzhugh.
\newblock Thresholds and plateaus in the hodgkin-huxley nerve equations.
\newblock {\em The Journal of general physiology}, 43(5):867--896, 1960.

\bibitem{2017Mathematics}
Fabrizio Gabbiani and Steven~James Cox.
\newblock Mathematics for neuroscientists.
\newblock {\em Academic Press}, 2017.

\bibitem{2000Isoform}
Junyuan Gao, X~Sun, V~Yatsula, RS~Wymore, and RT~Mathias.
\newblock Isoform-specific function and distribution of na/k pumps in the frog
  lens epithelium.
\newblock {\em The Journal of membrane biology}, 178(2):89--101, 2000.

\bibitem{2021Lymphatics}
Mariela Subileau and Daniel Vittet.
\newblock Lymphatics in eye fluid homeostasis: Minor contributors or
  significant actors?
\newblock {\em Multidisciplinary Digital Publishing Institute}, 10(7):582,
  2021.

\bibitem{uddin2022ocular}
Nasir Uddin and Matt Rutar.
\newblock Ocular lymphatic and glymphatic systems: implications for retinal
  health and disease.
\newblock {\em International Journal of Molecular Sciences}, 23(17):10139,
  2022.

\bibitem{2022Arterial}
Ravi~Teja Kedarasetti, Patrick~J. Drew, and Francesco Costanzo.
\newblock Arterial vasodilation drives convective fluid flow in the brain: a
  poroelastic model.
\newblock {\em Fluids and Barriers of the CNS}, 19(34), 2022.

\bibitem{rasmussen2018glymphatic}
Martin~Kaag Rasmussen, Humberto Mestre, and Maiken Nedergaard.
\newblock The glymphatic pathway in neurological disorders.
\newblock {\em The Lancet Neurology}, 17(11):1016--1024, 2018.

\bibitem{1995Generation}
A.~Amaratunga and R.~E. Fine.
\newblock Generation of amyloidogenic c-terminal fragments during rapid axonal
  transport in vivo of $\beta$-amyloid precursor protein in the optic nerve.
\newblock {\em Journal of Biological Chemistry}, 270(29):17268--17272, 1995.

\bibitem{1993Amyloid}
Peter~J. Morin, Carmela~R. Abraham, Anil Amaratunga, Robin~J. Johnson, Glenn
  Huber, Julie~H. Sandell, and Richard~E. Fine.
\newblock Amyloid precursor protein is synthesized by retinal ganglion cells,
  rapidly transported to the optic nerve plasma membrane and nerve terminals,
  and metabolized.
\newblock {\em Journal of Neurochemistry}, 61(2):662--669, 1993.

\bibitem{2021Perivascular}
K.~G. Freitas and A.~B. Leite.
\newblock Perivascular spaces and brain waste clearance systems: relevance for
  neurodegenerative and cerebrovascular pathology.
\newblock {\em Neuroradiology}, 63(10):1581--1597, 2021.

\bibitem{2022Electrostatic}
D.~W.~I. Smith and J.~A. Doe.
\newblock Electrostatic properties of amyloid $\beta$-protein and their impact
  on aggregation.
\newblock {\em Journal of Molecular Biology}, 432(4):789--798, 2022.

\bibitem{2021Optic}
Yi~Zhu, Shixin Xu, Robert~S. Eisenberg, and Huaxiong Huang.
\newblock Optic nerve microcirculation: Fluid flow and electro-diffusion.
\newblock {\em Physics of Fluids}, 33(4):041906, 2021.

\bibitem{hladky2022glymphatic}
Stephen~B Hladky and Margery~A Barrand.
\newblock The glymphatic hypothesis: the theory and the evidence.
\newblock {\em Fluids and Barriers of the CNS}, 19(1):9, 2022.

\bibitem{2020Circadian}
Lauren~M. Hablitz, Virginia Plá, Michael Giannetto, Hanna~S. Vinitsky, and
  Maiken Nedergaard.
\newblock Circadian control of brain glymphatic and lymphatic fluid flow.
\newblock {\em Nature Communications}, 11(1):4411, 2020.

\bibitem{2024Abrain}
Keelin Quirk, Kimberly A.~S. Boster, and Jeffrey Tithof.
\newblock A brain-wide solute transport model of the glymphatic system.
\newblock {\em Journal of the Royal Society Interface}, (TN.219):21, 2024.

\end{thebibliography}

	\newpage
	\appendix
	
	\begin{figure}
		\centering
		\includegraphics[width=1\linewidth]{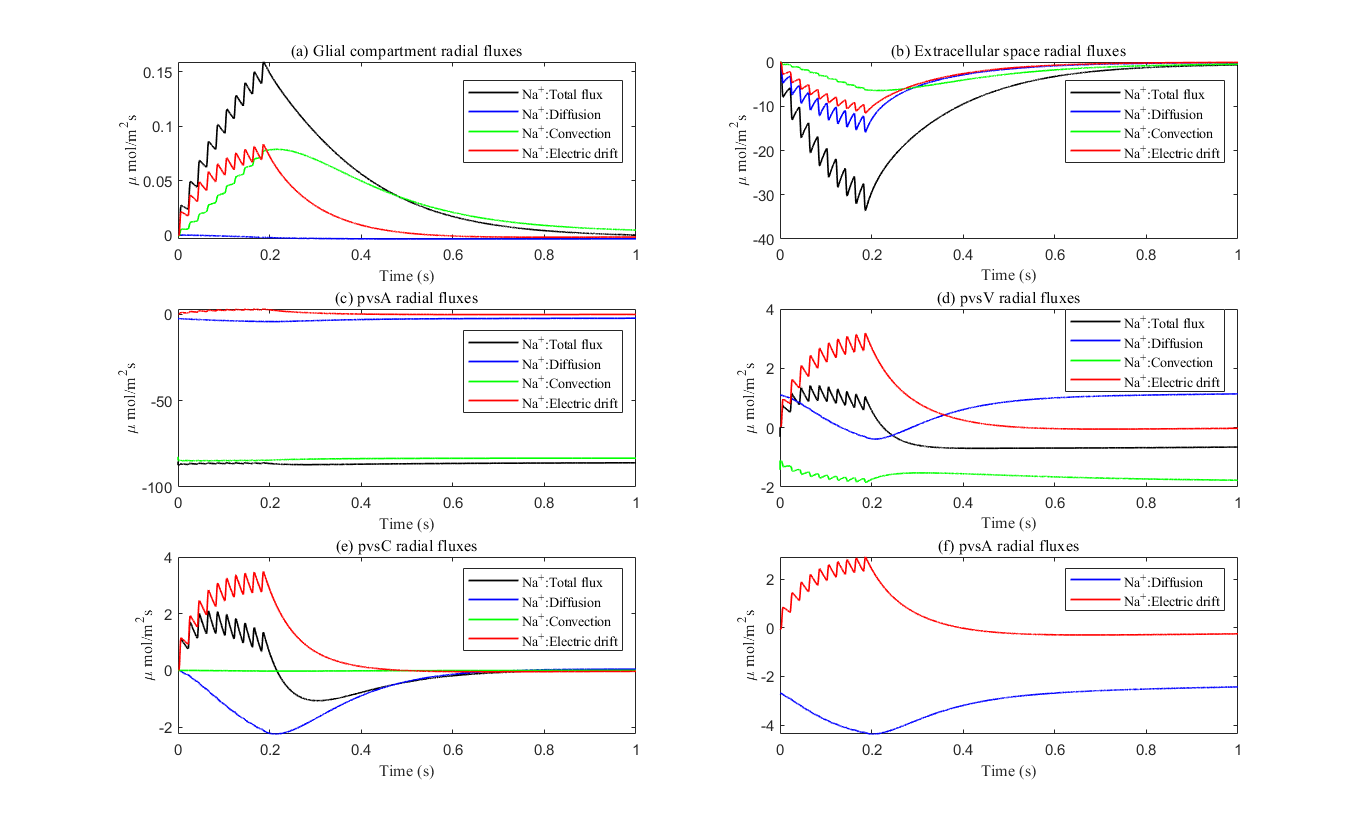}
		\caption{\label{fig:NaDiffusion} Average radial direction sodium fluxes components in the intradomain.}
	\end{figure}
	
\end{document}